\documentclass[10pt,twocolumn]{article}
\usepackage{amsmath}
\usepackage{amsfonts}
\usepackage{amssymb}
\usepackage{graphicx}
\usepackage{grffile}
\usepackage{braket}
\usepackage{bbold}
\usepackage{subcaption}
\usepackage{stfloats}
\usepackage{listings}
\usepackage{bm}
\usepackage{qcircuit}
\usepackage{xcolor}
\usepackage{authblk}
\usepackage[hidelinks]{hyperref}
\usepackage[percent]{overpic}

\title{Natural orbitals and sparsity of quantum mutual information}
 
\author[1,2]{Leonardo Ratini}
\author[2]{Chiara Capecci}
\author[1]{Leonardo Guidoni
\thanks{leonardo.guidoni@univaq.it}}

\affil[1]{Dipartimento di Scienze Fisiche e Chimiche, Universit\`a degli Studi dell'Aquila, Coppito, L'Aquila, Italy}
\affil[2]{Dipartimento di Ingegneria e Scienze dell'Informazione e Matematica, Universit\`a degli Studi dell'Aquila, Coppito, L'Aquila, Italy}

\date{\today}

\begin{document}

\twocolumn[
\begin{@twocolumnfalse}

\maketitle
\begin{abstract} 
Natural orbitals, defined in electronic structure and quantum chemistry as the (molecular) orbitals diagonalizing the one-particle reduced density matrix of the ground state, have been conjectured for decades to be the perfect reference orbitals to describe electron correlation. In the present work 
we applied the Wavefunction-Adapted Hamiltonian Through Orbital Rotation (WAHTOR) method to study correlated empirical ans\"atze for quantum computing.
In all representative molecules considered, we show that the converged orbitals are coinciding with natural orbitals. 
Interestingly, the resulting quantum mutual information matrix built on such orbitals is also maximally sparse, providing a clear picture that such orbital choice is indeed able to provide the optimal basis to describe electron correlation. The correlation is therefore encoded in a smaller number of qubit pairs contributing to the quantum mutual information matrix.  

\end{abstract}
\vspace{1cm}
\end{@twocolumnfalse}]

\section{INTRODUCTION}
Electronic structure theory and its implementation into algorithms and computer programs allow the study of many physical and chemical processes with both theoretical and application interests. In cases where electron correlation plays an important role, high-level classical electronic structure and quantum chemistry tools are necessary, albeit they become limited by the bad scaling of the required computational time with the size of the system. Quantum devices may offer new opportunities to tackle these problems, in principle leading to exponential advantage. \cite{Kitaev1995, Abrams1997, Abrams1999, Nielsen2010} 
Many algorithms have been developed, \cite{Bauman2021, Barison2021, McArdle2019} including hybrid quantum-classical algorithms that use classical and quantum resources in combination, exploiting at best the features of both devices.

Among the hybrid algorithms, the variational ones are widespread in different fields of application such as machine learning \cite{Cilimberto2018, Dunjko2020}, combinatorial optimization \cite{Nannicini2019,Egger2021}, strongly correlated systems \cite{Huggins2020} and quantum chemistry \cite{OMalley2016,Bauer2020}.
One of the most widely used algorithms is the variational quantum eigensolver (VQE) \cite{Fedorov2022,Cao2019,Peruzzo2014}, which has the advantage to be suited for the implementation on noisy intermediate-scale quantum (NISQ) devices \cite{Corcoles2020, Preskill2018, Moll2018}. The electronic structure problem is tackled by writing the molecular Hamiltonian in the second quantisation scheme and mapping fermions into qubits using one of the many encoding methods available \cite{Jordan1928,Bravyi2002,McArdle2020}.
In this scheme, the Hamiltonian is written as a weighted sum of strings of Pauli operators. Variational parameters of wave function ans\"atze are therefore optimised using a classical computer whereas the quantum processor is used to calculate expected values of the Hamiltonian.

Variational wave functions, i.e. variational quantum circuits, used in VQE algorithm are often inspired by existing quantum chemistry methods such as unitary Coupled Cluster (UCC) \cite{Lee2019, Romero2019}. 
These classes of Quantum-Chemistry-inspired ans\"atze have the advantage of being directly derived by a consolidated theoretical frameworks, but they have the drawback of being realised by circuits composed by a large number of quantum gates.
For instance in the case of UCCSD the number of gates scales as $\mathcal{O}\left((M - N )^2 N^2 \right)$, where $M$ represents the number of spin-orbitals and $N$ the number of electrons \cite{Lee2019}.
On NISQ devices, which are limited by a small number of gates due to the short coherence time of the qubits, heuristic ans\"atze are rather considered, since they can better exploit the capabilities of the hardware, at the cost of loosing the chemical/physical meaning of their construction \cite{Kandala2017}.
Many groups have recently developed different variants of the VQE algorithm, with the aim to improve its efficiency, enhance the quality of the wave function, and reduce the circuit depth \cite{Grimsley2019, Benfenati2021, Stair2021, Ganzhorn2019, Egger2023, Tkachenko2020, Meitei2021}. 

In this paper, we analyse the mutual information matrix of shallow depth ans\"atze in which we optimised both the variational parameters and the orbitals used to construct the ans\"atze. To achieve this goal we used a recently optimised version of the Wavefunction-Adapted Hamiltonian Through Orbital Rotation (WAHTOR) algorithm \cite{Ratini2022, Ratini2023}. 
The WAHTOR algorithm was introduced in order to have a good description of the system considered using a short circuit, improving the energy results without increasing the number of the quantum gates in the circuit. The WAHTOR algorithm finds the optimal unitary transformation of the Hartree-Fock orbitals for a given ansatz of a fixed topology. In this perspective, the Hamiltonian is optimised (or, in another perspective, orbitals are optimised) in order to adapt to the shape of the ansatz. From the computational point of view, this optimization is achieved by using the gradient of the energy with respect the parameters of the unitary transformation, without significant overload. In an improved version of the algorithm \cite{Ratini2023}, we exploited higher-order derivatives, reformulating the algorithm for a more efficient and faster convergence.

By analysing the resulting optimised orbitals we have observed that they are converging very closely to the natural orbitals basis \cite{Loewdin1955,Davidson1972,Szabo2012}, namely the orbitals determined by the unitary transformation that diagonalizes the one-particle reduced density matrix of that ground state. The WAHTOR procedure seems therefore to lead to the finding of ``maximally correlated orbitals'' which were indeed identified as natural orbitals along the years.
In addition, anticipating our results, we have discovered that in this WAHTOR-optimised/natural orbital basis the sparsity of the mutual information matrix is indeed minimal, as expected when the basis set better express the electronic correlation of the system.

In section \ref{computational_details} we shortly review the improved WAHTOR algorithm,  we show the computational details of the algorithm, the systems studied, the ans\"atze used and the analysis performed. Section \ref{results} show how, for the systems of study, the optimized basis sets converge to these of natural orbitals. Finally, in last section we focus on the key results obtained and related discussions.\\

\section{COMPUTATIONAL DETAILS}
\label{computational_details}

{\it \textbf{WAHTOR algorithm}}\\
The main idea of the WAHTOR algorithm is to adapt the Hamiltonian to a given empirical quantum circuit \cite{Ratini2022}. In this regard we exploit the invariance of the Hamiltonian eigenvalues with respect unitary transformations. 
In particular, the Hamiltonian, in the second quantization formalism, is the following:
\begin{equation}
H=\sum_{i,j}h_{ij}a^\dag_i a_j+\frac{1}{2}\sum_{c,d,e,f}g_{cdef}a^\dag_c a^\dag_d a_e a_f 
\label{eq:H}
\end{equation}
where the sums runs on $M$, the number of spin-orbitals and $h_{ij}$ and $g_{cdef}$ are respectively the one-body and two-body integrals.
We choose the following particular form for the unitary operator $\hat{U}$: 
\begin{equation}
   \hat{U}=\hat{U}(\bm{R})=e^{\sum_{jk}(i\bm{R}\cdot\bm{T})_{jk}a_j^{\dag}a_k}
\end{equation}
where the components of $\bm{T}$ are the generators of the unitary Lie group with dimension $MxM$, where $M$ is the number of spin-orbitals. The elements of the group are obtained by linearly combining the generators with real coefficients, represented by the components of the vector $\bm{R}$. The operator $\hat{U}^{\dag}\hat{H}\hat{U}$ is the Hamiltonian $H$ expressed in a new basis set. Given a parametrized quantum state $\ket{\Psi(\bm{\theta})}$ on a quantum circuit, we can define the energy cost function 
\begin{equation}
E(\bm{R},\bm{\theta})=\bra{\Psi(\bm{\theta})}H(\bm{R})\ket{\Psi(\bm{\theta})}
\label{eq:energy}
\end{equation}

Both sets of parameters can be alternatively optimized: the $\bm{\theta}$ parameters with a VQE algorithm and the $\bm{R}$ parameters with one of the procedures detailed in the non-adiabatic version of the WAHTOR algorithm \cite{Ratini2023}. When convergence is reached, we obtain the converged set of spin-orbitals.
\\
{\it \textbf{Simulated systems}}\\
We considered seven different molecular systems: $H_2$, $LiH$, $HF$, $BeH_2$, $H_2O$, $H_2S$, and $NH_3$. Depending on the systems we used different basis sets and frozen core approximations, as detailed by table \ref{table_molecule_info}.
\begin{table*}[!ht]
\caption{Geometric distances between atoms, basis set, number of qubits and groups of orbitals with same geometrical symmetries for each molecule under investigation.}
\label{table_molecule_info}
\begin{center}
\newcommand{\minitab}[2][l]{\begin{tabular}#1 #2\end{tabular}}
\begin{tabular}{| c | c | c | c | c | c |}
\hline
\hline 
molecule &  atomic positions (\AA{}) &  basis set &  qubits & orbitals groups & \# of generators\\ 
\hline
\hline
$H_2$   & \minitab[c]{H 0.0 0.0 0.0 \\ H 0.0 0.0 0.74}  & 6-31g  & 8 & \minitab[c]{\{1,3\},\{2,4\}} & 2\\
\hline
$LiH$   & \minitab[c]{Li 0.0 0.0 0.0 \\ H 0.0 0.0 1.595}  & sto-3g  & 10 & \minitab[c]{\{1,2,5\}} & 3\\
\hline
$HF$    & \minitab[c]{F 0.0 0.0 0.0 \\ H 0.0 0.0 0.917}  & sto-3g  & 10 & \minitab[c]{\{1,2,5\}} & 3\\
\hline 
$BeH_2$ & \minitab[c]{H 0.0 0.0 0.0 \\ Be 0.0 0.0 1.334 \\ H 0.0 0.0 2.668}  & sto-3g  & 12 & \minitab[c]{\{1,5\},\{2,6\}} & 2\\
\hline 
$H_2O$  & \minitab[c]{O 0.0 0.0 0.0 \\ H 0.757 0.586 0.0 \\ H -0.757 0.586 0.0}
& sto-3g  & 12 & \minitab[c]{\{1,3,5\},\{2,6\}} & 4\\
\hline
$H_2S$  & \minitab[c]{S 0.0 0.0 0.0 \\ H 0.0 0.9616 -0.9269 \\ H 0.0 -0.9616 -0.9269}  & sto-3g  & 12 & \minitab[c]{\{1,3,6\},\{2,5\}} & 4\\
\hline        
$NH_3$  & \minitab[c]{N 0.0 0.0 0.1211 \\ H 0.0 0.9306 -0.2826 \\ H 0.8059 -0.4653 -0.2826	\\ H -0.8059 -0.4653 -0.2826}  & sto-3g  & 14 & \minitab[c]{\{1,4,5\},\{2,6\},\{3,7\}} & 5\\
\hline		
\end{tabular}
\end{center}
\end{table*}

We used the Jordan-Wigner encoding method \cite{Jordan1928} obtaining a number of qubits ranging from 8 to 14. The Hartree-Fock orbitals and the terms of the molecular Hamiltonian have been obtained using the PySCF package \cite{pyscf}. The WAHTOR algorithm has been implemented by a in-house produced Python program exploiting the Qiskit libraries \cite{qiskit}.
Both the VQE and the WAHTOR algorithm were performed for each molecule and the energy results obtained were compared taking into account the fraction of correlation energy $\epsilon$ , defined by:
\begin{equation}
\epsilon =\frac{E_{calculated}-E_{HF}}{E_{full-CI}-E_{HF}}
\end{equation}
where $E_{full-CI}$ is the exact energy, $E_{calculated}$ is the energy calculated using VQE or WAHTOR algorithm and $E_{HF}$ is the Hartree-Fock energy.
The percentage of correlation energy values are shown in the table \ref{table_molecule_results} for each molecule under investigation.
\\

\begin{figure}[!h]
\centering
\begin{equation}
    \Qcircuit @C=0.7em @R=0.3em {
   	    	& \lstick{\ket{0}} & \gate{R_y(\theta_0)} & \ctrl{1} & \qw & \qw & \qw & \qw & \qw & \qw & \gate{R_y(\theta_8)} & \qw &\\
   	    	& \lstick{\ket{1}} & \gate{R_y(\theta_1)} & \targ & \ctrl{1} & \qw & \qw & \qw & \qw & \qw & \gate{R_y(\theta_9)} & \qw &\\
   	    	& \lstick{\ket{2}} & \gate{R_y(\theta_2)} & \qw & \targ & \ctrl{1} & \qw & \qw & \qw & \qw & \gate{R_y(\theta_{10})} &\qw &\\
   	    	& \lstick{\ket{3}} & \gate{R_y(\theta_3)} & \qw & \qw & \targ & \ctrl{1} & \qw & \qw & \qw & \gate{R_y(\theta_{11})} & \qw &\\
   	    	& \lstick{\ket{4}} & \gate{R_y(\theta_4)} & \qw & \qw & \qw & \targ & \ctrl{1} & \qw & \qw & \gate{R_y(\theta_{12})} &\qw & \\
   	    	& \lstick{\ket{5}} & \gate{R_y(\theta_5)} & \qw & \qw & \qw & \qw & \targ & \ctrl{1} & \qw & \gate{R_y(\theta_{13})} & \qw & \\
   	    	& \lstick{\ket{6}} & \gate{R_y(\theta_6)} & \qw & \qw & \qw & \qw & \qw & \targ & \ctrl{1} & \gate{R_y(\theta_{14})} & \qw &\\
   	    	& \lstick{\ket{7}} & \gate{R_y(\theta_7)} & \qw & \qw & \qw & \qw & \qw & \qw & \targ & \gate{R_y(\theta_{15})} & \qw & \\
   	    	\ar@{-}[]+<5.3em,15.8em>;[]+<5.3em,-0.0em>
   	    	\ar@{-}[]+<5.4em,15.8em>;[]+<5.4em,-0.0em>
   		    \ar@{-}[]+<20.5em,15.8em>;[]+<20.5em,-0.0em>
   		    \ar@{-}[]+<20.6em,15.8em>;[]+<20.6em,-0.0em>
        }
\notag
\end{equation}
\caption{Ladder entangler map for an 8-qubits system. The quantum gates between the two barriers compose the block (identified by bold lines in the figure). The block is repeated a certain number of times $d$, which defines the depth of the circuit.}
\label{fig:ansatz_example}
\end{figure}
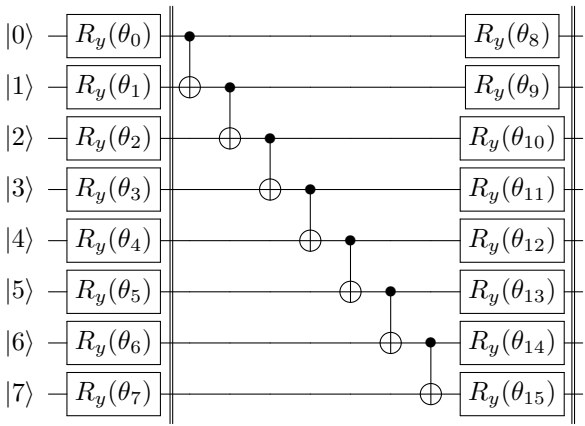

{\it \textbf{Implementation of quantum algorithm}}\\
The parametrized quantum circuit chosen in the present work belongs to the so-called heuristic ansatzes and it is built as follows. Firstly, a repetition of single qubit rotations around the y-axis is applied to the n-qubits Hartree-Fock reference state. After such rotations, an entangling block composed of CNOT gates followed by another set of independent rotations on each qubit has been applied. The block composed by CNOT gates and rotations can be therefore repeated a certain number of times, each time using different variational parameters, as shown in figure \ref{fig:ansatz_example}. The total number of blocks determines the circuit depth. In this work, we considered depth 2 for diatomic molecules, depth 4 for molecules composed of 3 atoms and depth 6 for the $NH_3$ system. In all systems the CNOT gates are arranged to form a ladder, as shown in figure \ref{fig:ansatz_example}.
 
For each molecular system, we executed 100 VQE runs starting from random variational parameters and the lowest energy result was considered as starting point for the WAHTOR algorithm. We optimized the ansatz parameters using the L-BFGS-B optimizer \cite{Byrd1995}. Whereas the Hamiltonian optimization step is performed by applying the non-adiabatic trust region optimizer defined in ref. \cite{Ratini2023}.
The convergence is reached when in two successive VQE energies in the WAHTOR algorithm differ less than $10^{-6}$ Hartree. All  simulations are executed using the statevector simulator. 
\\

{\it \textbf{Hamiltonian parameters}}\\
For the implementation of the WAHTOR algorithm, the matrices that determine the basis change take spin symmetry into account, i.e. the spin-up and the spin-down spin-orbitals are never combined linearly during the optimization. In this way, as generators for a system with $m=M/2$ orbitals, we have a set of $\frac{m^2+m}{2}$ symmetric and $\frac{m^2-m}{2}$ antisymmetric matrices. 
Since the Hamiltonian and the wavefunction are reals  we take into account only the generators giving rise to real unitary matrices.
Moreover, we restrict the optimization to combine linearly only orbitals having the same molecular symmetries. The list of orbitals with having the same symmetries and reported, for each molecule, in the last but one column of table \ref{table_molecule_info}.
These two considerations reduce the number of generators for the unitary group that satisfy these constraints, reducing the components of the $\bm{T}$ vector as well as the components of $\bm{R}$, the number of optimized parameters.
As a consequence, the computational cost is reduced and the number of generators for each system is reported in the last column of table \ref{table_molecule_info}.
\\

{\it \textbf{Quantum mutual information matrix}}\\
Since we use Jordan-Wigner encoding method correlations between qubits is reflected into the correlation between the occupation numbers. A useful measure of correlation between the qubit pairs is represented by the quantum mutual information \cite{Zeng2019}.
Considering a system of $n$ qubits, each one defined in its own Hilbert space $\mathcal{H}_i$ for $i=1,\dots,n$, the state of the composite system $\ket{\Psi}$ belongs to the space composed by the tensor product of the individual spaces $\ket{\Psi} \in \bigotimes_{i=1}^n \mathcal{H}_i$. The density operator has the following expression:
\begin{equation}
\rho= \ket{\Psi}\bra{\Psi}
\end{equation}
By defining the set of the indices $\Omega=\{1,\dots,n\}$ we can select a set of indices $k\in \Omega$.
The reduced density matrix on qubits with indices in $k$ is obtained from the density matrix $\rho$ tracing out over all the indices with the exception of $k$:
\begin{equation}
\rho_{k}=Tr_{\{\Omega-k\}}(\rho)
\end{equation}
In particular, we define the one-qubit and two-qubits reduced density matrices as:
\begin{align}
& \rho_{i}=Tr_{\{\Omega-i\}}(\rho)\\
& \rho_{ij}=Tr_{\{\Omega-(i,j)\}}(\rho)
\label{eq:rdm}
\end{align}
where $i$ and $j$ refer to the qubits considered.
Now, the calculation of Von Neumann entropy $S$ 
\begin{equation}
S(\rho)= -Tr(\rho \ln \rho)
\end{equation}
for the reduced density matrices defined in equations \ref{eq:rdm} can be obtained as follows:
\begin{align}
    & S(\rho_i)=-Tr(\rho_i \ln \rho_i) \\
    & S(\rho_{ij})=-Tr(\rho_{ij}ln\rho_{ij})
\end{align}
Finally, the quantum mutual information between qubits $i$ and $j$ is defined as:
\begin{equation}
I_{ij}=\left(S(\rho_{i})+S(\rho_{j})-S(\rho_{ij})\right)(1-\delta_{ij})
\label{mutual_info}
\end{equation}
We represent the quantum mutual information as a symmetric matrix with elements along the diagonal equal to zero by definition, following reference \cite{Tkachenko2020}. The value of the correlation between the qubits can be observed in the upper or the lower triangular part of this matrix.
\\

{\it \textbf{Orbital analysis}}\\
In order to carry out an analysis on the convergence of the orbitals obtained with the WAHTOR algorithm to the natural ones, as conjectured, the dimensionless parameter $\delta\in [0,1]$ has been defined.
It measures how much the converged WAHTOR orbitals ($\ket{W}$) are close to the natural ones ($\ket{NO}$), calculated from the exact Full-CI state. \cite{Szabo2012} The definition of this parameter is the following:
\begin{equation}
\delta=\frac{\sum_i(|\braket{W_i|NO_i}|-|\braket{HF_i|NO_i}|}{N-\sum_i|\braket{HF_i|NO_i}|}
\label{delta}
\end{equation}
where $N$ is the number of orbitals considered and $\ket{HF}$ are the canonical Hartree-Fock ones. \cite{Helgaker2000,Giuliani2005} So the closer the $\delta$ value is to 1, the more the optimized orbitals are similar to the reference natural orbitals (corresponding to $\delta=1$); the closer it is to zero, the more the optimized orbitals are closer to the Hartree-Fock orbitals (corresponding to $\delta=0$).

\begin{table*}
\caption{VQE and WAHTOR percentage correlation energy results and geometric distance from the reference natural orbitals for each molecule under investigation.}
\label{table_molecule_results}
\centering
\newcommand{\minitab}[2][l]{\begin{tabular}#1 #2\end{tabular}}
\begin{tabular}{| c | c | c | c | c | c | c |}
\hline
\hline 
molecule & FCI ($E_h$) & VQE ($E_h$) &\% VQE & WAHTOR ($E_h$) & \% WAHTOR &  $\delta$\\ 
\hline
\hline
$H_2$ & -1.866777 & -1.848151 & 25.25 & -1.861338 & 78.17  & 0.997  \\
\hline
$LiH$ & -1.079201 & -1.073586 & 72.14 & -1.078143 & 94.75  & 0.999  \\
\hline
$HF$ & -27.985031 & -27.979893 & 80.11  & -27.983232 & 93.03  & 0.999  \\
\hline 
$BeH_2$ & -3.940331 & -3.921635 & 46.36  & -3.921642 & 46.38  & $\ket{HF}\approx\ket{NO}$  \\
\hline 
$BeH_2$ & -3.940331 & -3.914403 & 25.61  & -3.914778 & 26.68  & $\ket{HF}\approx\ket{NO}$  \\
\hline 
$BeH_2$ & -3.940331 & -3.911841 & 18.25  & -3.911863 & 18.31  & $\ket{HF}\approx\ket{NO}$  \\
\hline 
$H_2O$ & -23.544497 & -23.520888 & 52.22 & -23.531435 & 73.56 & 0.971  \\
\hline
$H_2S$ & -15.971142 & -15.949235 & 48.34 & -15.953189  & 57.66 & 0.888  \\
\hline        
$NH_3$ & -20.100795 & -20.060577 & 38.87 & -20.060781 & 39.18 & 0.999\\
\hline		
\end{tabular}
\end{table*}
\section{RESULTS}
\label{results}
\begin{figure*}
\caption{Representation of the $H_2O$ molecular orbitals in the frozen core approximation optimized by the algorithm. Each column corresponds to a different basis set: the first one represents the Hartree-Fock orbitals, the second one the WAHTOR-optimized orbitals and the one the natural orbitals.\\}
    \centering
    $\ket{HF}_1$
    \begin{overpic}[width=0.25\textwidth]{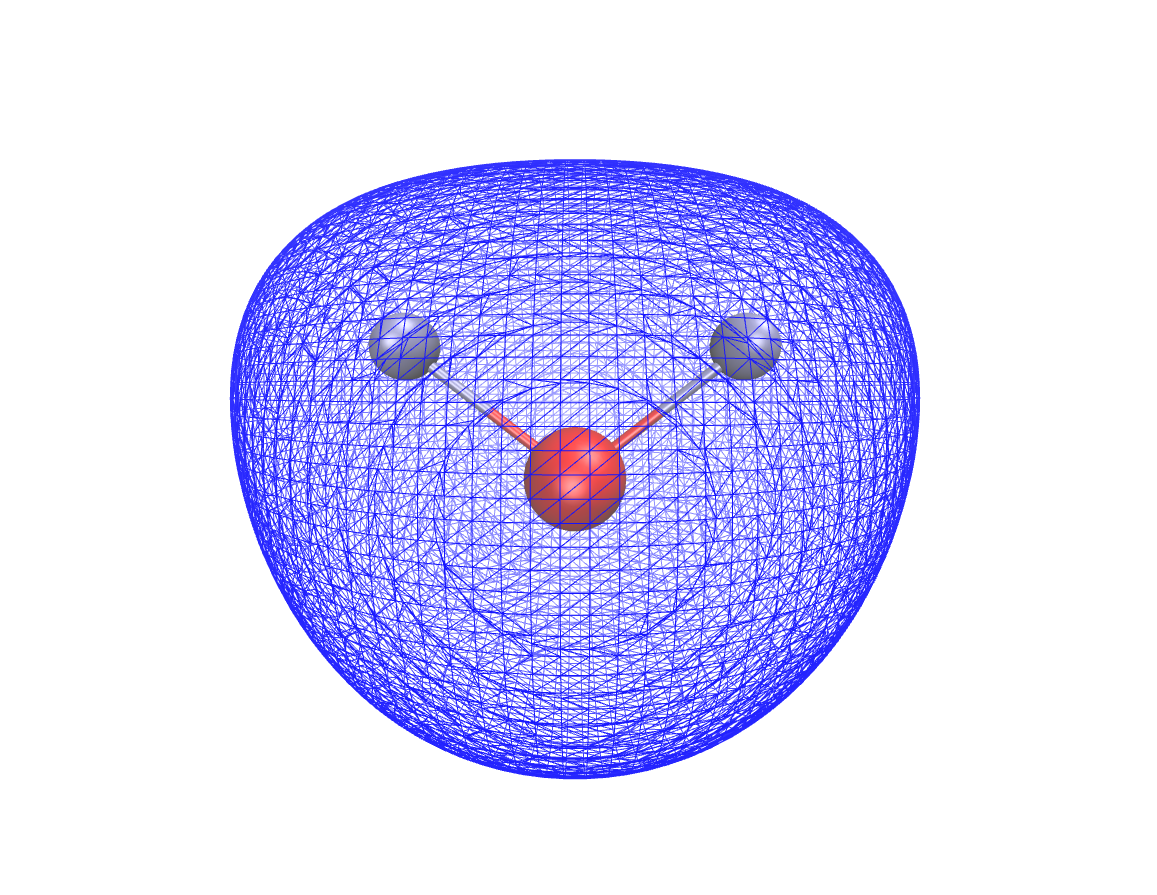}\hfill
    \put (30,80) {\huge$\ket{HF}$}
    \end{overpic}
    $\ket{W}_1$
    \begin{overpic}[width=0.25\textwidth]{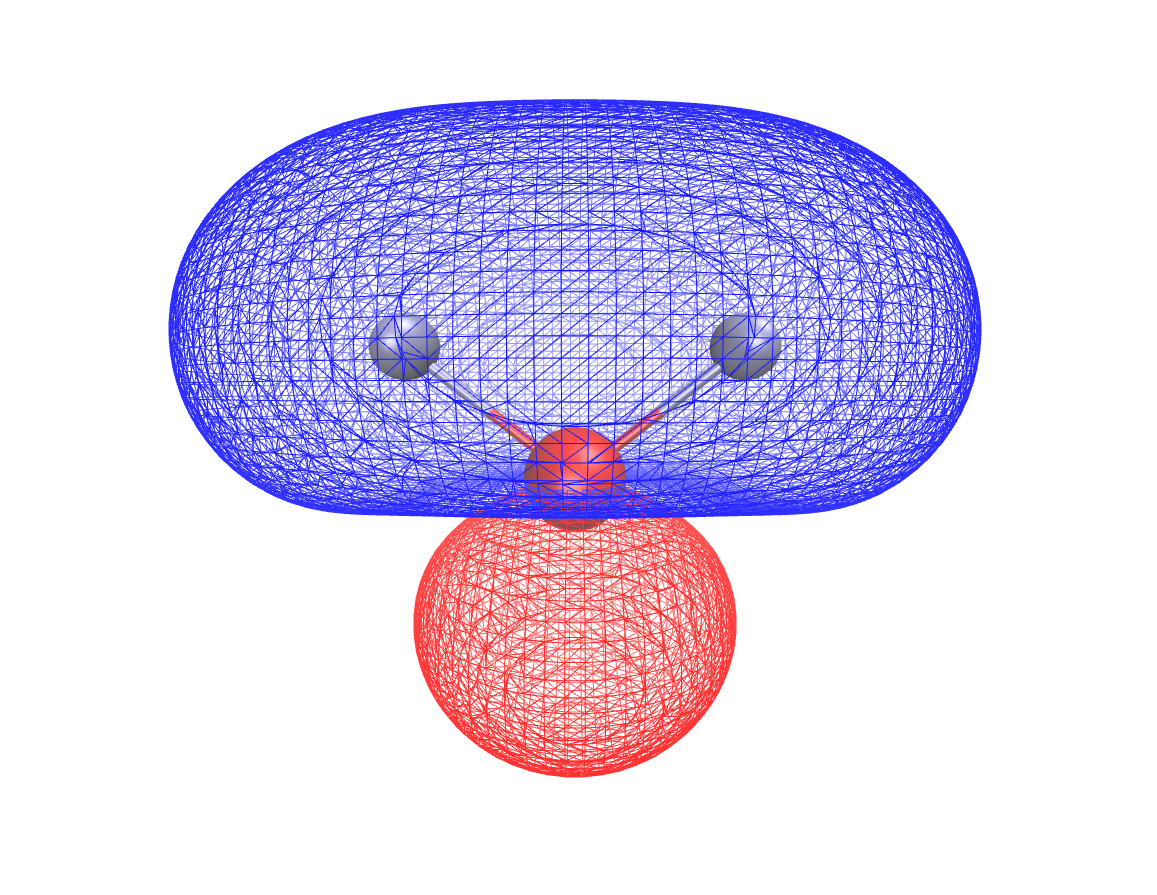}\hfill
    \put (35,80) {\huge$\ket{W}$}
    \end{overpic}
    $\ket{NO}_1$
    \begin{overpic}[width=0.25\textwidth]{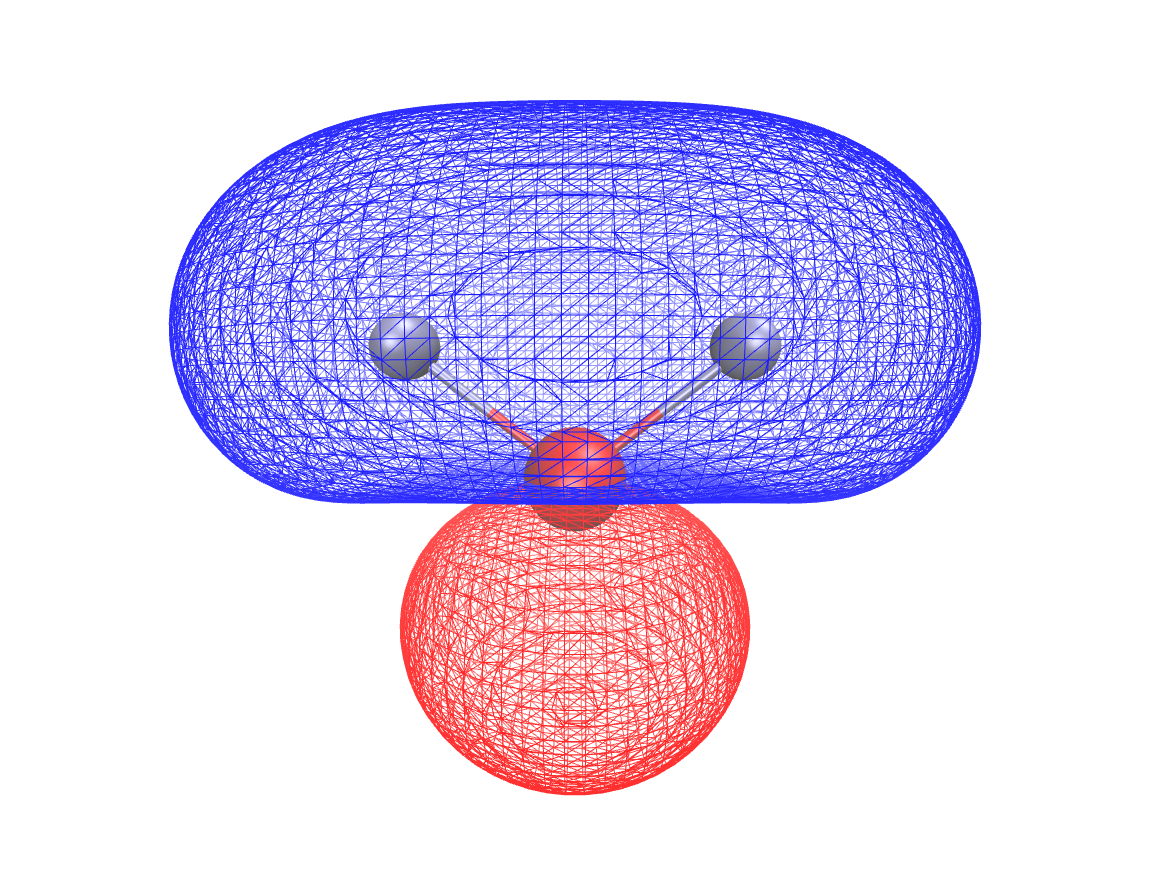}
    \put (30,80) {\huge$\ket{NO}$}
    \end{overpic}
    \\[\smallskipamount]
    $\ket{HF}_2$  \includegraphics[width=.25\textwidth]{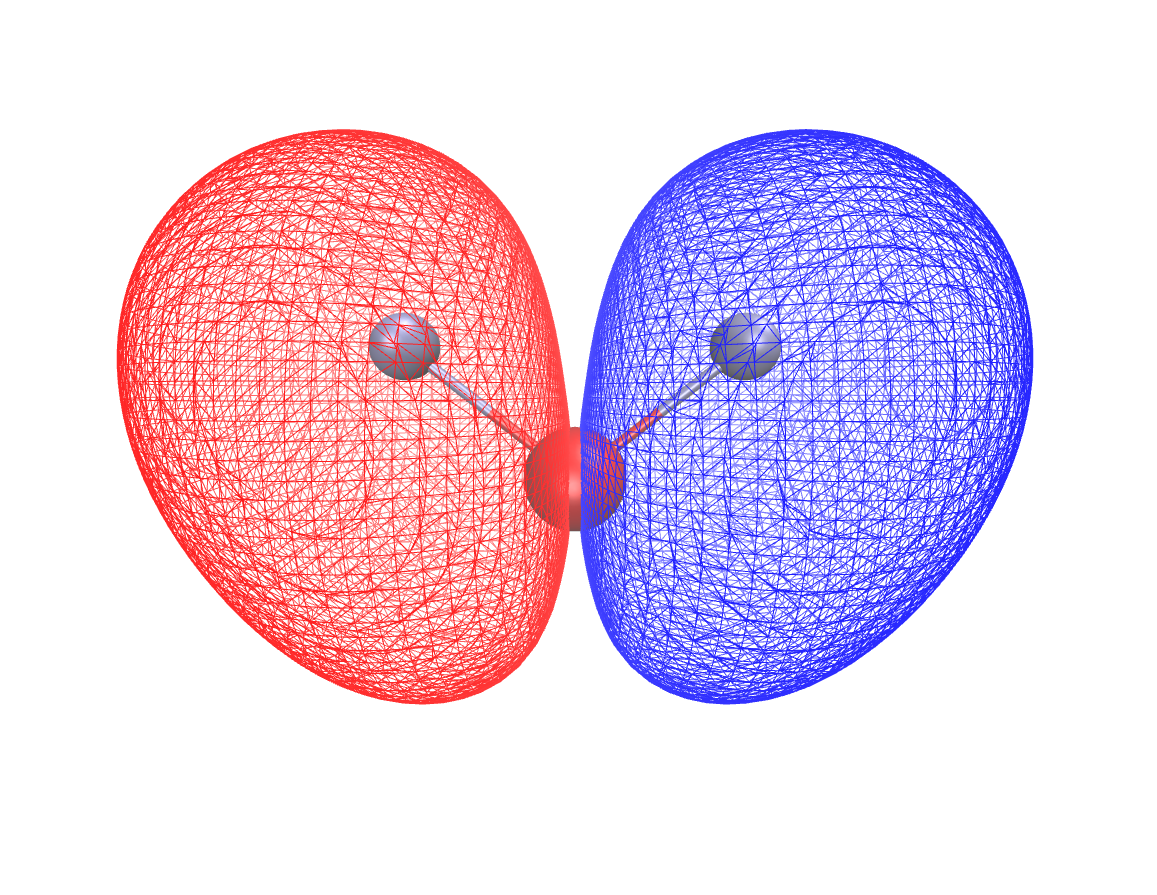}\hfill
    $\ket{W}_2$
    \includegraphics[width=.25\textwidth]{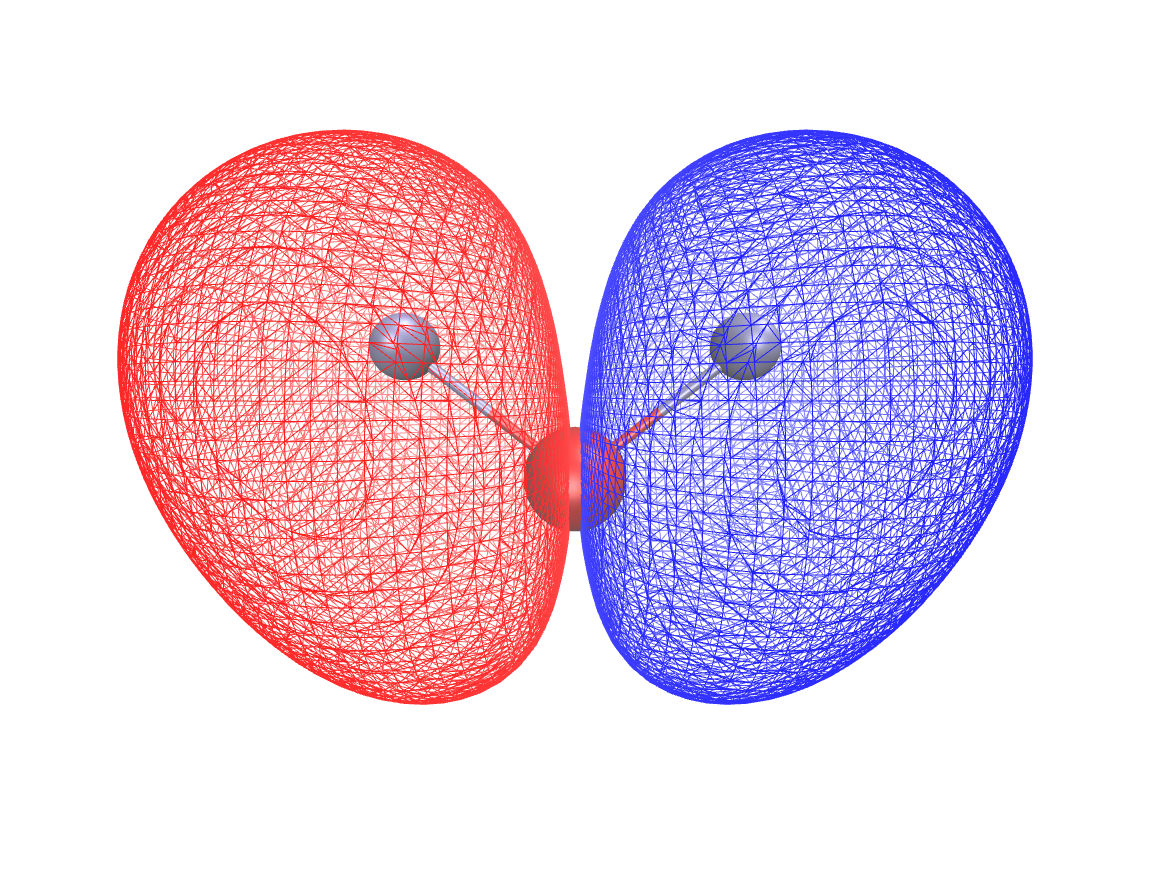}\hfill
    $\ket{NO}_2$
    \includegraphics[width=.25\textwidth]{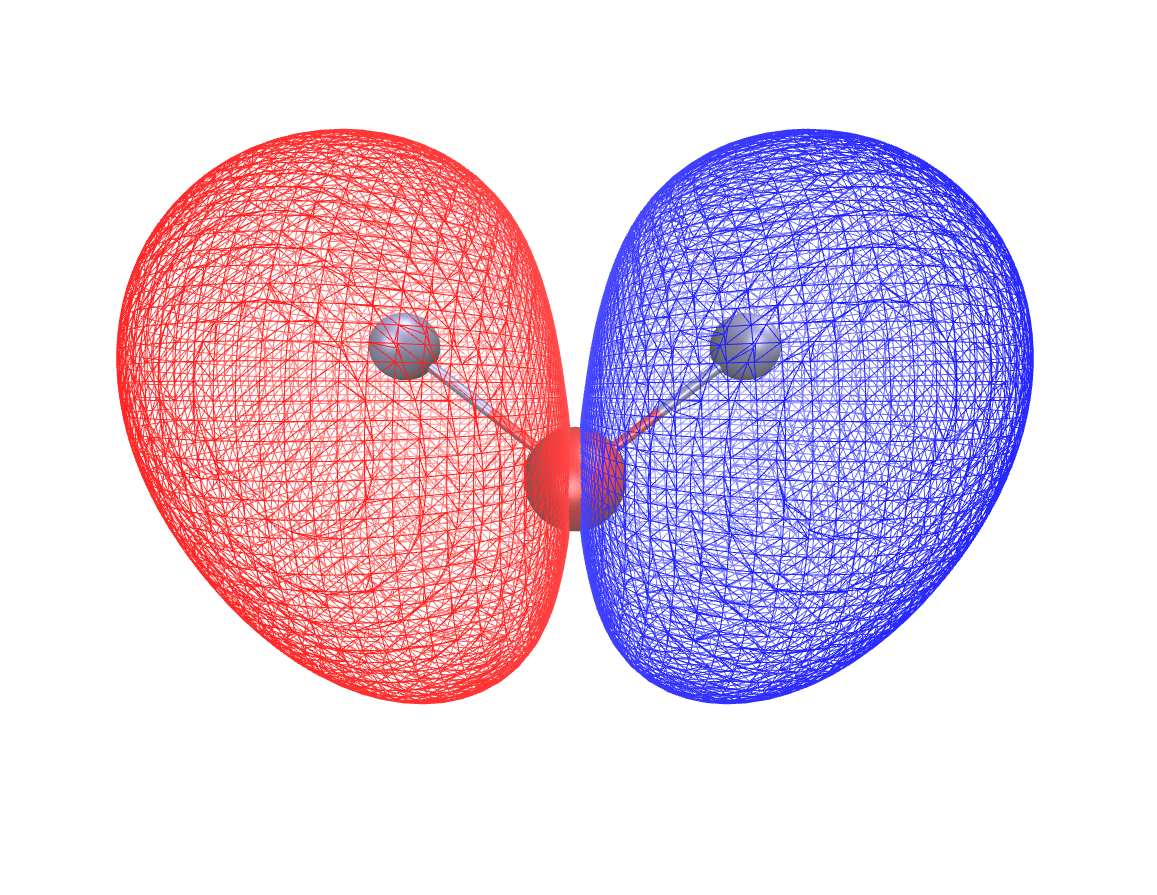}
    \\[\smallskipamount]
    $\ket{HF}_4$
    \includegraphics[width=.25\textwidth]{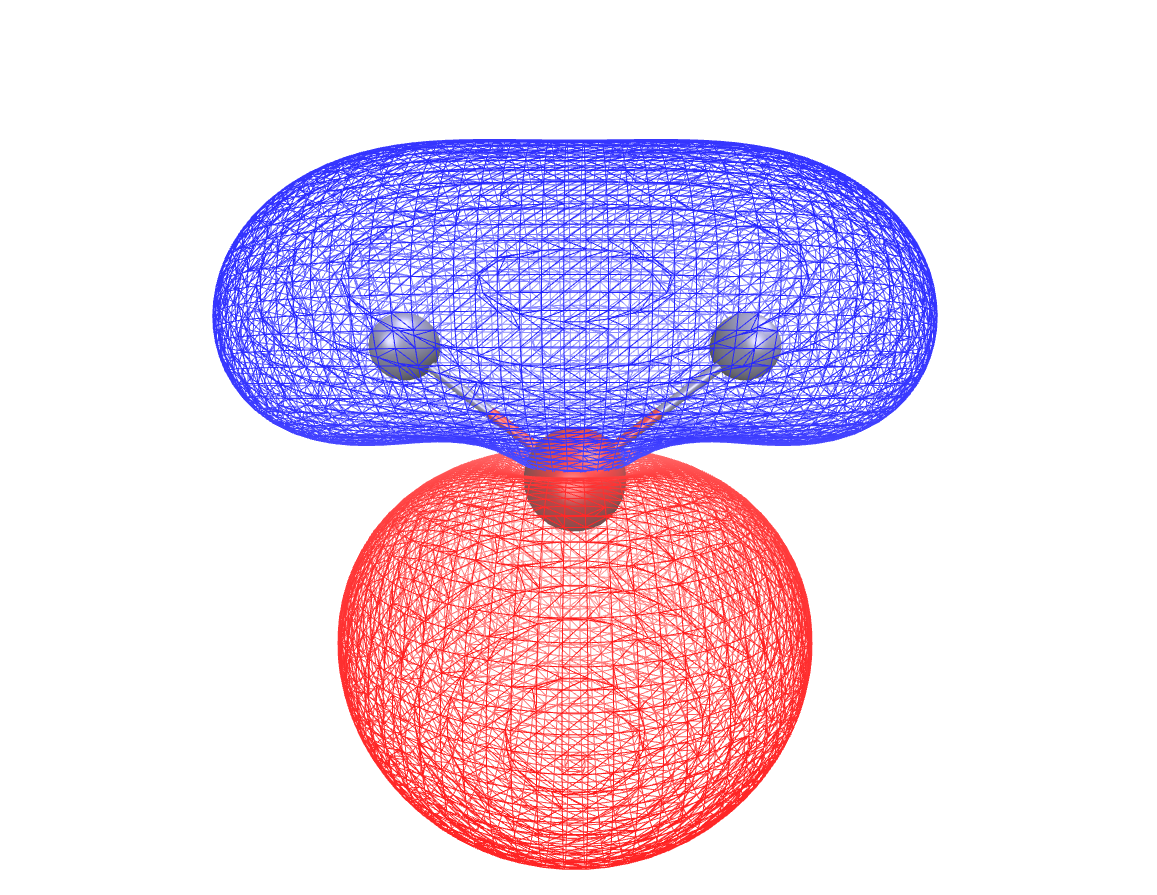}\hfill
    $\ket{W}_4$
    \includegraphics[width=.25\textwidth]{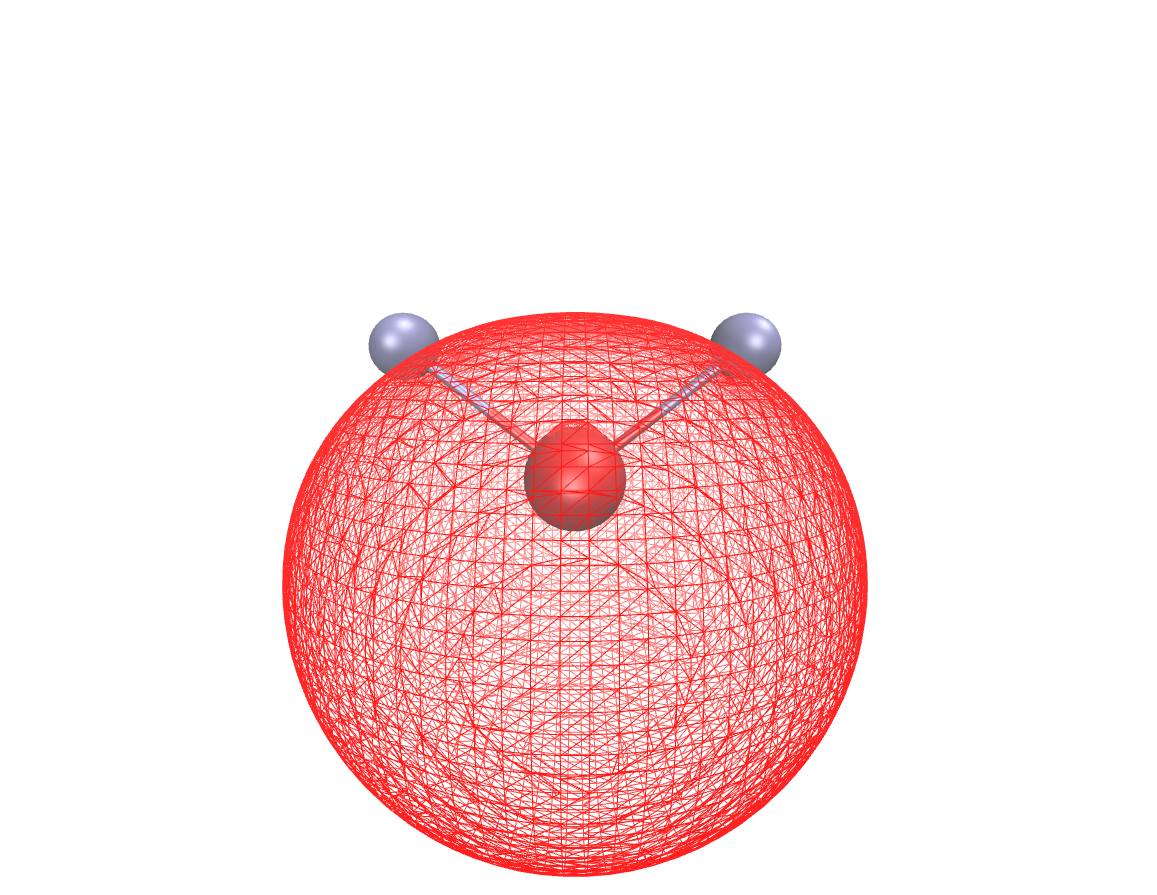}\hfill
    $\ket{NO}_4$
    \includegraphics[width=.25\textwidth]{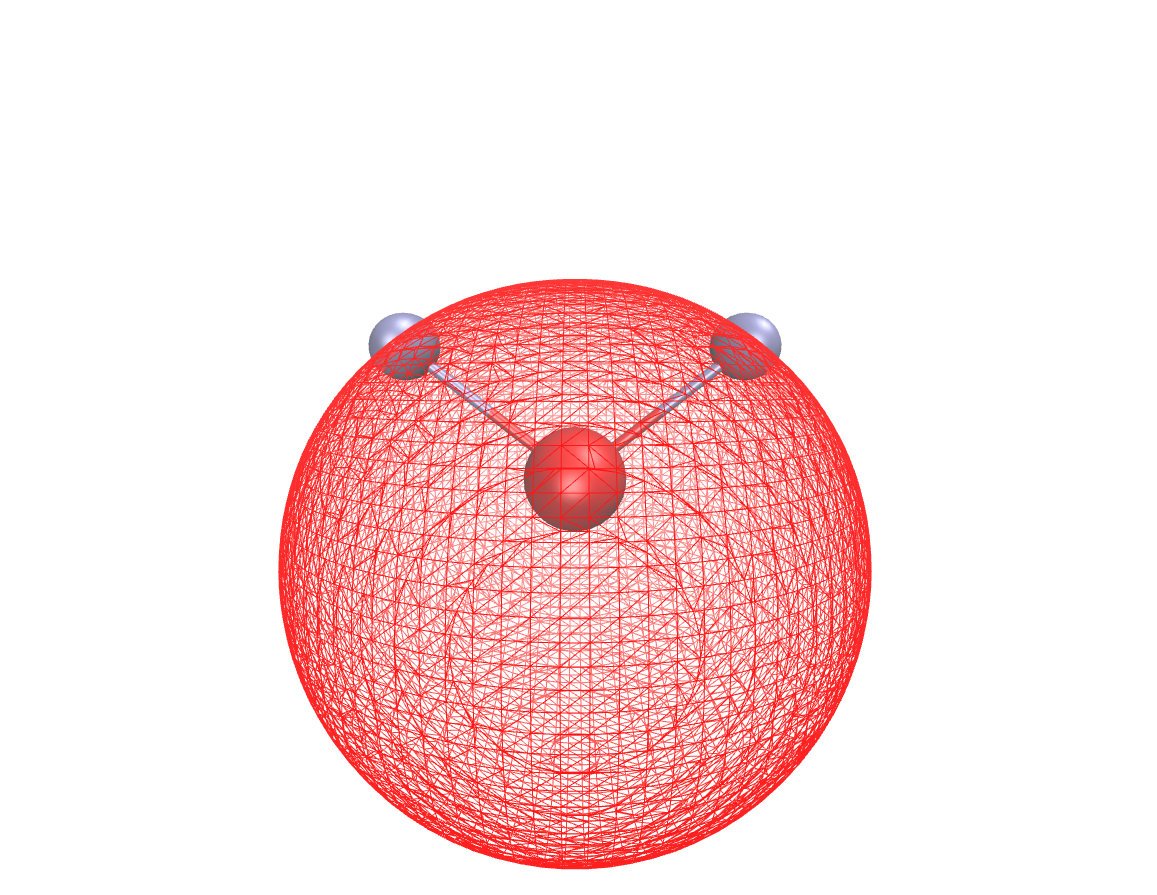}
    \\[\smallskipamount]
    $\ket{HF}_5$
    \includegraphics[width=.25\textwidth]{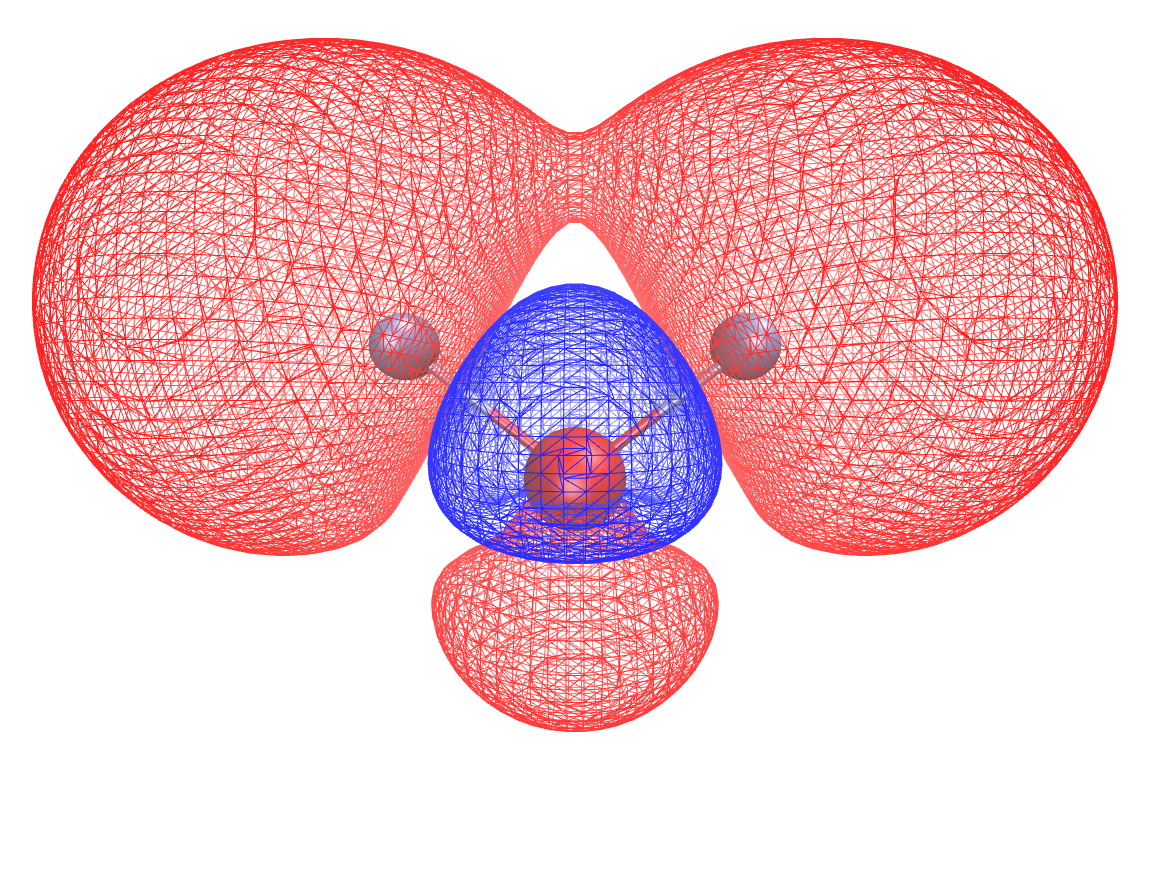}\hfill
    $\ket{W}_5$
    \includegraphics[width=.25\textwidth]{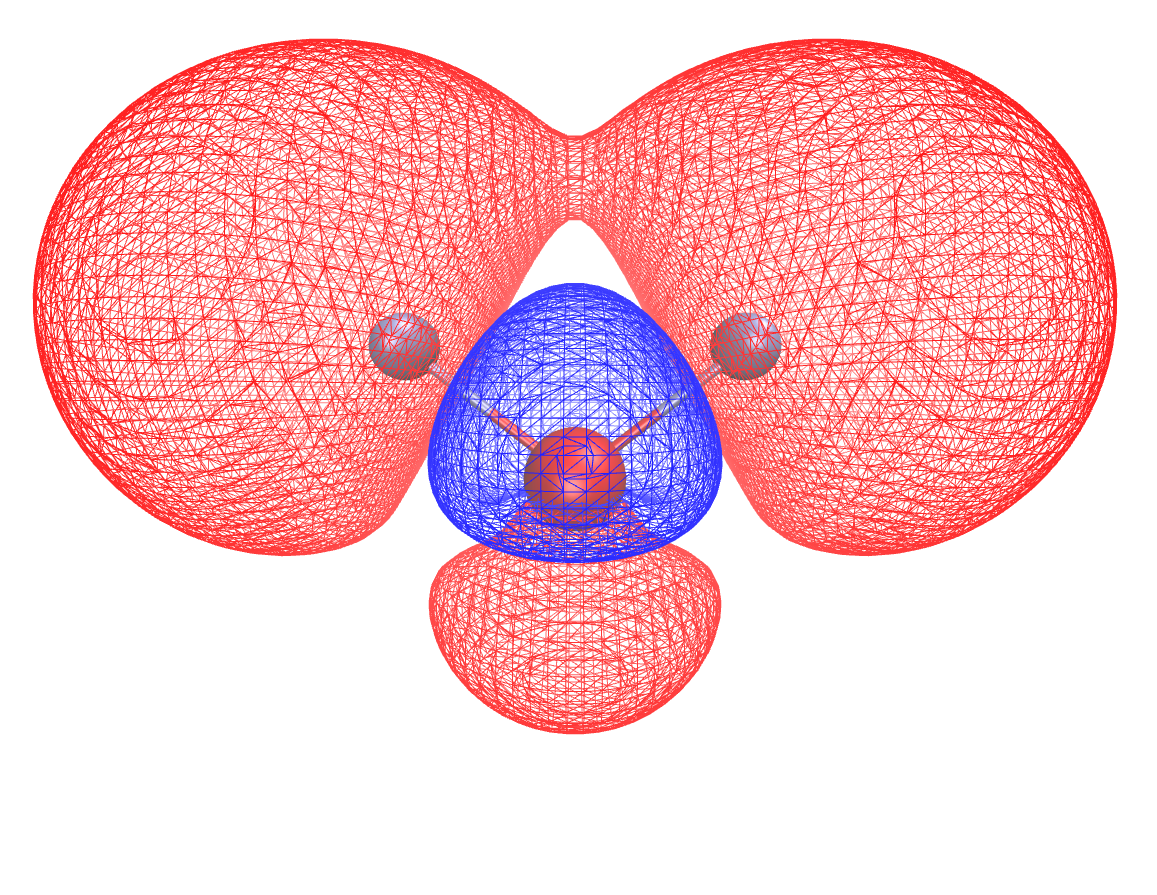}\hfill
    $\ket{NO}_5$
    \includegraphics[width=.25\textwidth]{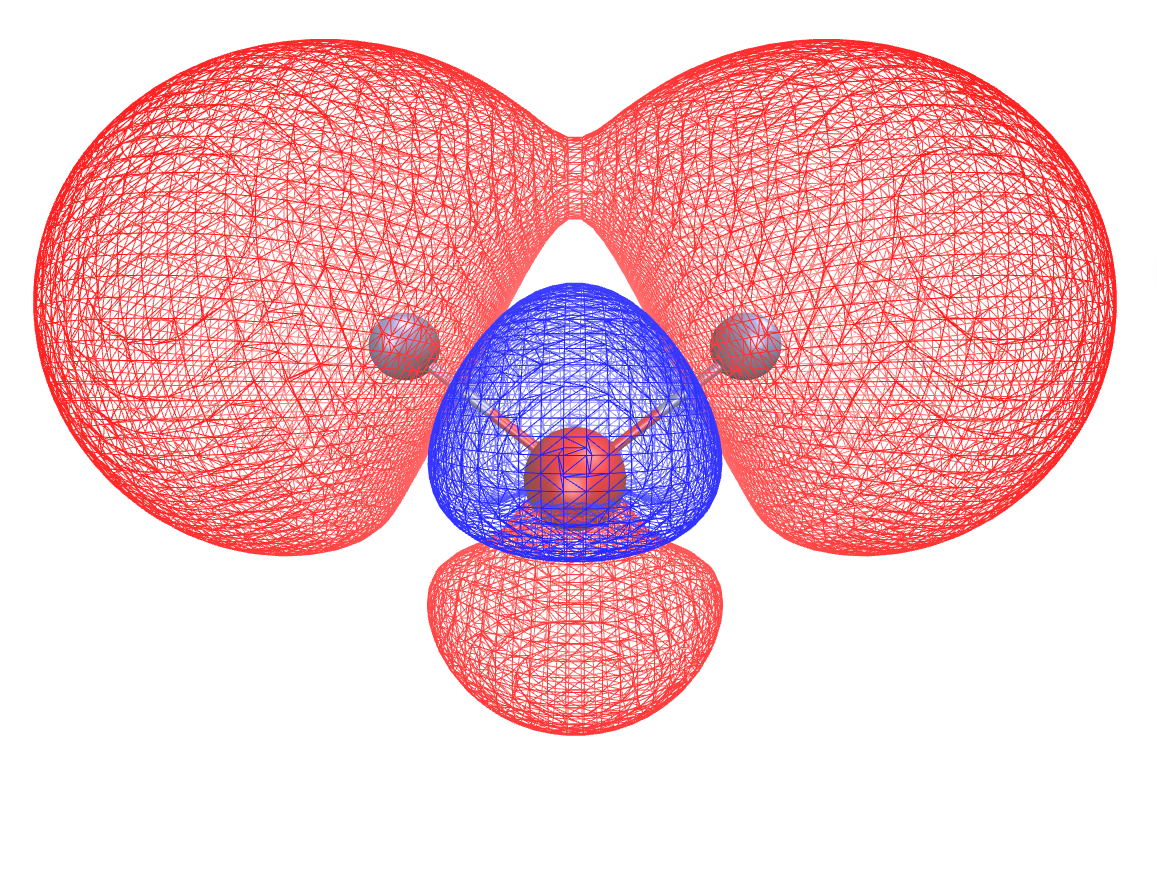}
    \\[\smallskipamount]
    $\ket{HF}_6$
    \includegraphics[width=.25\textwidth]{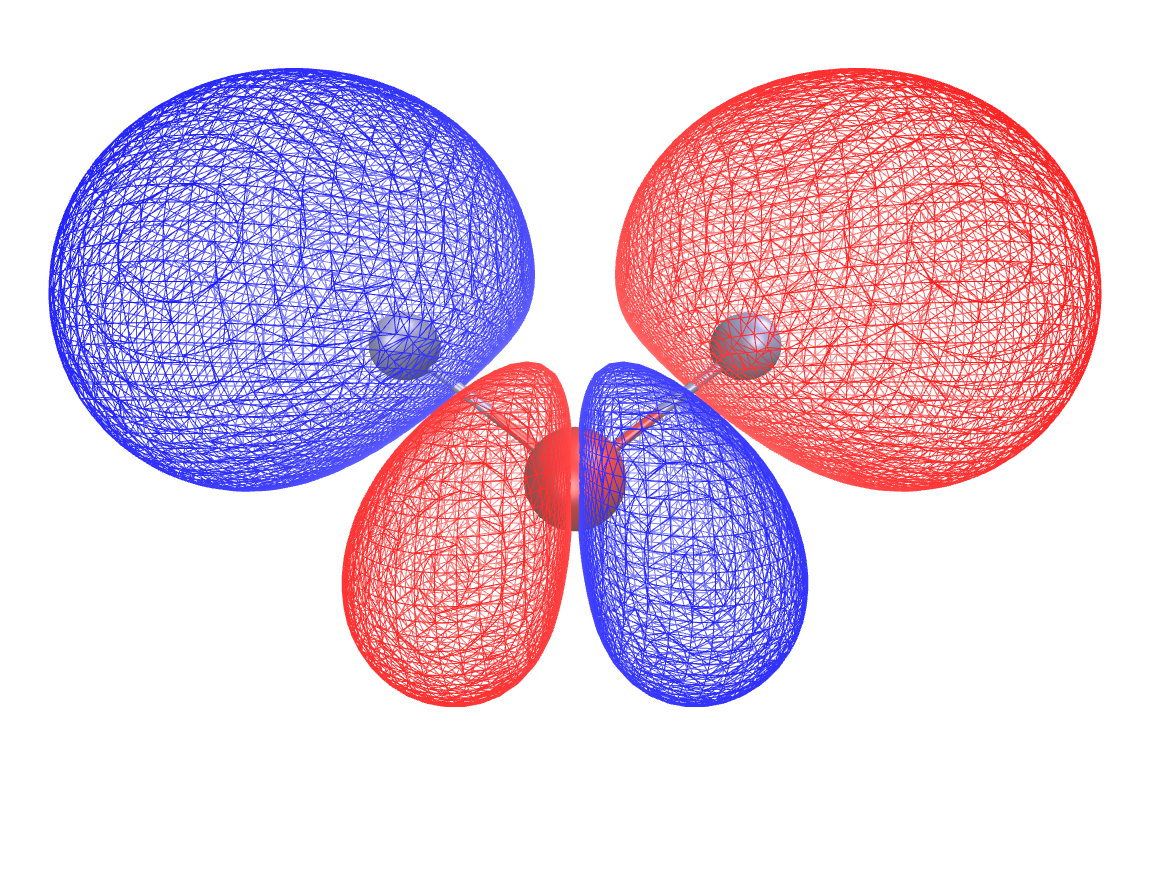}\hfill
    $\ket{W}_6$
    \includegraphics[width=.25\textwidth]{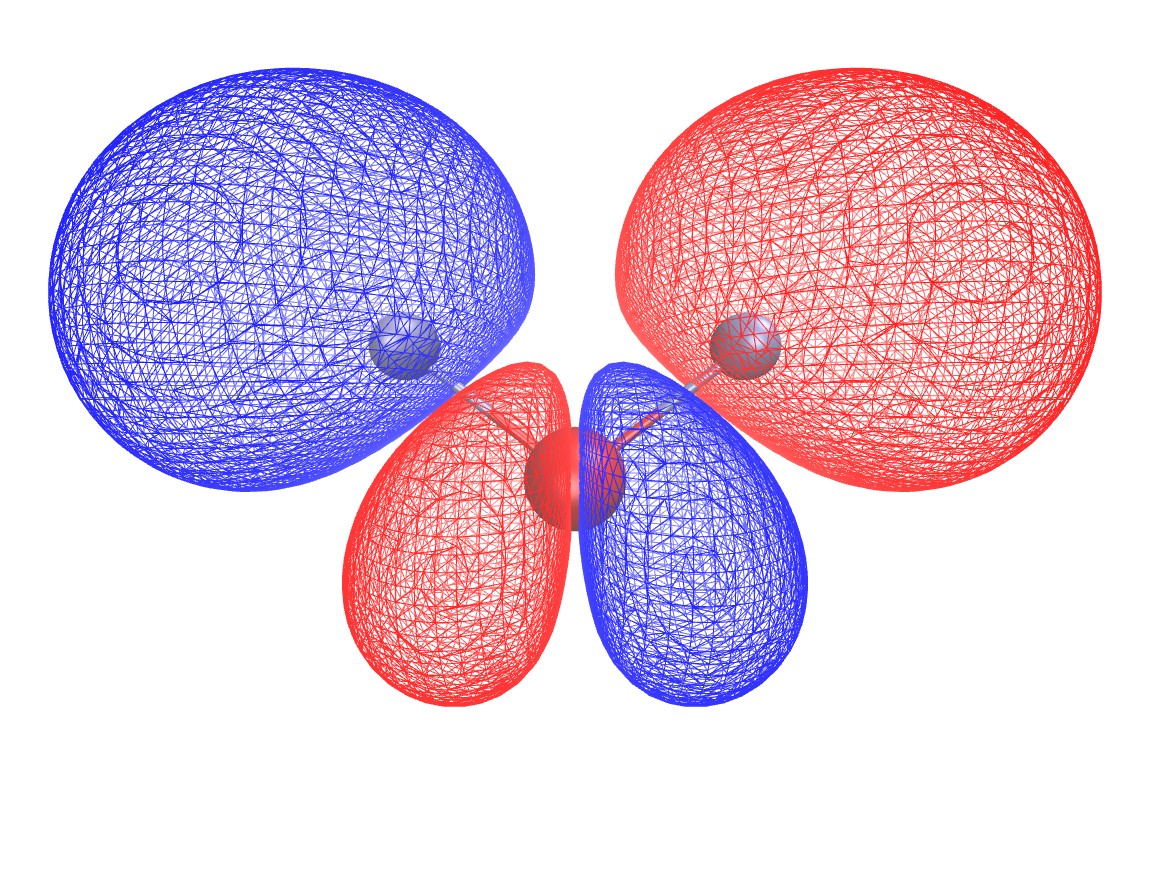}\hfill
    $\ket{NO}_6$
    \includegraphics[width=.25\textwidth]{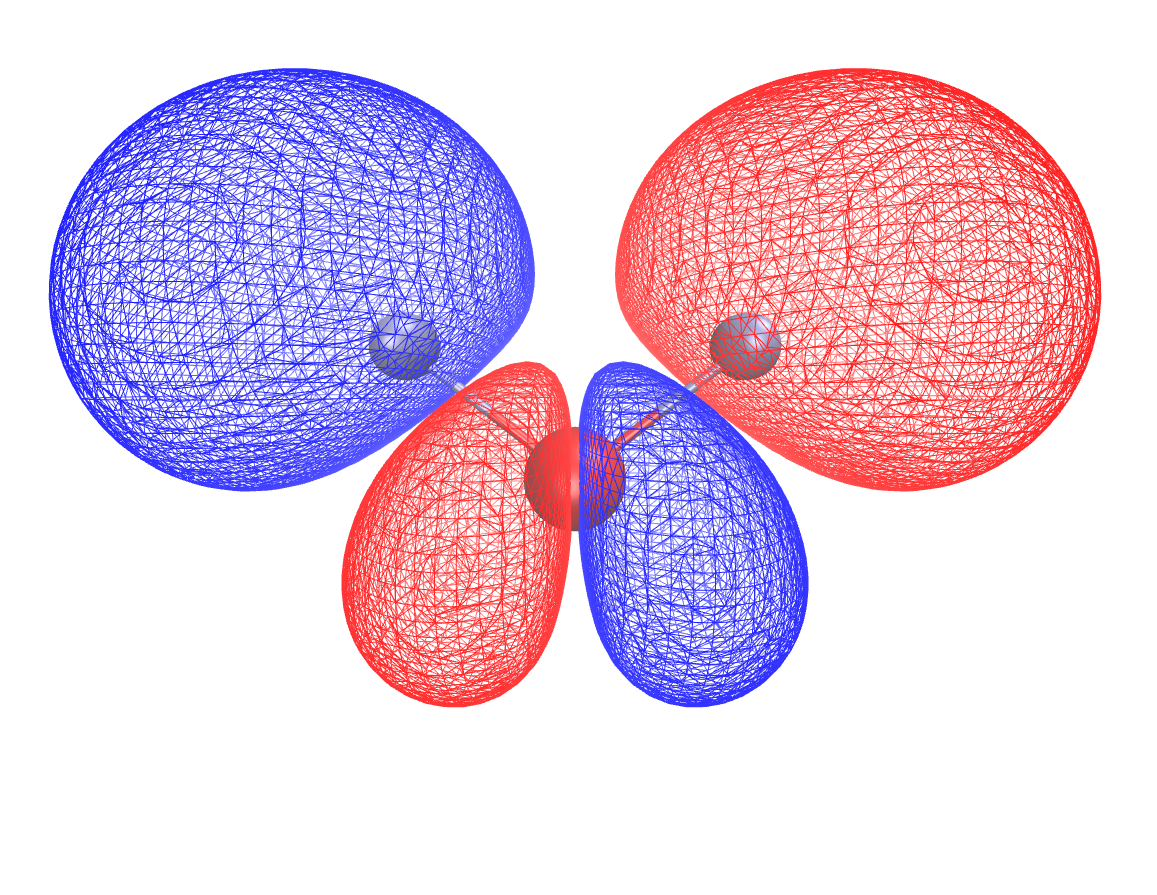}  
    
\label{fig:orbitals}
\end{figure*}

In this section we show the results obtained for the seven molecules taken into consideration: for each studied system the energy results obtained with the VQE and the WAHTOR algorithms have been compared, as well as the convergence of the single-particle basis set to the natural orbitals.
The results are summarised in table \ref{table_molecule_results}: the first column of the table reports the molecule under consideration while the second column is the reference energy expressed in Hartree ($E_h$) calculated using the Full Configuration Interaction (FCI) method. In the third and fourth columns, the lowest energy values obtained from 100 VQE runs and the related percentage correlation energies are reported, respectively. The fifth and sixth columns show the converged energies of the WAHTOR algorithm and the corresponding percentages of correlation energies.
We note that, for each molecule, our method always improves the VQE results. Interestingly, in the case of $BeH_2$ the orbital optimization does not change significantly the energy value for all the three VQE results considered, the reasons for this pitfall will be clearer after the orbital analysis and we will come back to this point later in the text.  
In the last column of table \ref{table_molecule_results} we reported the $\delta$ parameter, as defined by equation \ref{delta} in the computational details. This parameter is a measure of the distance between optimised orbitals and natural orbitals, ranging from $\delta=0$ (Hartree-Fock orbitals) to $\delta=1$ (natural orbitals of the groundstate calculated with FCI). 
Focusing on the $H_2O$ molecule, we note that $\delta=0.971$ so the optimized single-particle basis set corresponds approximately to the natural orbitals.
The similarity is evident as it can be also shown by comparing HF canonica orbitals, WAHTOR-optimised orbitals and natural orbitals reported respectively in the first, second and third columns of figure \ref{fig:orbitals}. 
Each row in the figure shows the five orbitals that are optimized by our algorithm. The orbitals in the first column have been obtained with the restricted Hartree-Fock method using the frozen core approximation. We split the orbitals into two groups that contain the ones with the same geometric symmetries: A1, A3 and A4 belong to the first group while A2 and A5 to the second one.
As explained in section \ref{computational_details}, two orbitals are combined linearly only if they belong to the same group so, since the missing orbital in the figure \ref{fig:orbitals} has a unique geometric symmetry, it has not been optimized by the algorithm and therefore has not been reported.
The WAHTOR-optimised orbitals in the second column of figure \ref{fig:orbitals} show that only the orbitals A1 and A3 have been appreciably modified becoming B1 and B3. By comparing the second and third columns of the figure, we infer that the WAHTOR-optimised and the natural orbitals are almost identical.
The same conclusions can be inferred by taking into account the other molecular systems reported in table \ref{table_molecule_results}. Indeed, also in these cases, the $\delta$ parameter is close to one, almost independently by the energy reached by the optimization algorithm.
An exception is made by the $BeH_2$ molecule for which the WAHTOR algorithm does not optimize the VQE energies, as shown in the last column of table \ref{table_molecule_results}. To emphasize this result, we considered the best three different VQE energies and tried to optimize the cost function without significant improvements. In terms of our previous statements, this is not surprising if we consider that the natural orbitals and the Hartree-Fock canonical orbitals are almost identical for this system.

In summary, our WAHTOR procedure is leading to orbitals which are extremely close to natural orbitals. We can therefore make the conjecture that our algorithm is guiding the Hamiltonian (i.e. rotating the orbitals) in such a way they are getting closer to the natural orbitals of the groundstate expressed in the Hartree-Fock basis set, which in principle can be obtained only by diagonalizing the one-body density matrix of this state.

\begin{figure*}
\centering
\subfloat[VQE state in HF canonical orbitals]
   {\includegraphics[width=0.42\textwidth]{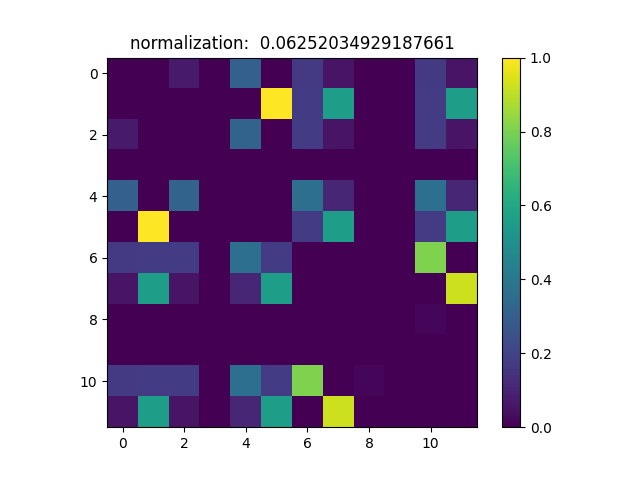}} \quad
\subfloat[WAHTOR state in WAHTOR-optimized orbitals]
   {\includegraphics[width=0.42\textwidth]{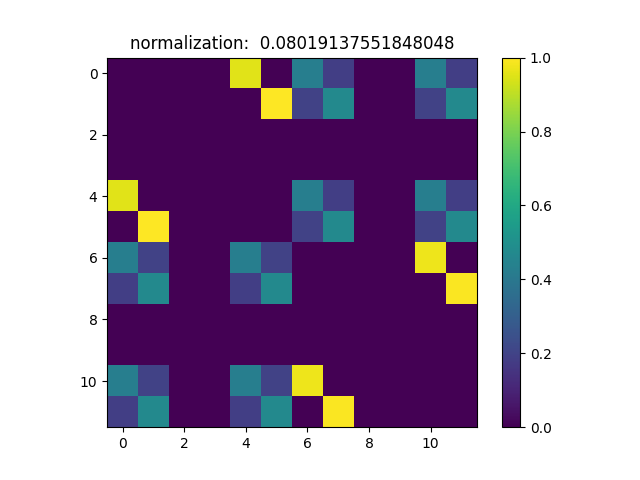}} \\
\subfloat[Groundstate in HF canonical orbitals]
   {\includegraphics[width=0.42\textwidth]{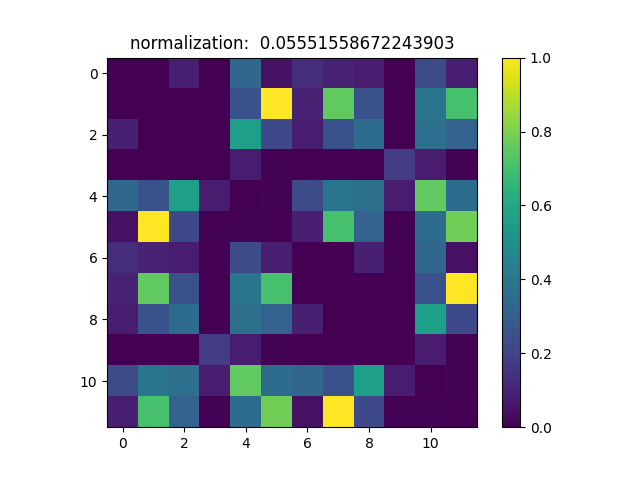}} \quad
\subfloat[Groundstate in WAHTOR-optimized orbitals]
   {\includegraphics[width=0.42\textwidth]{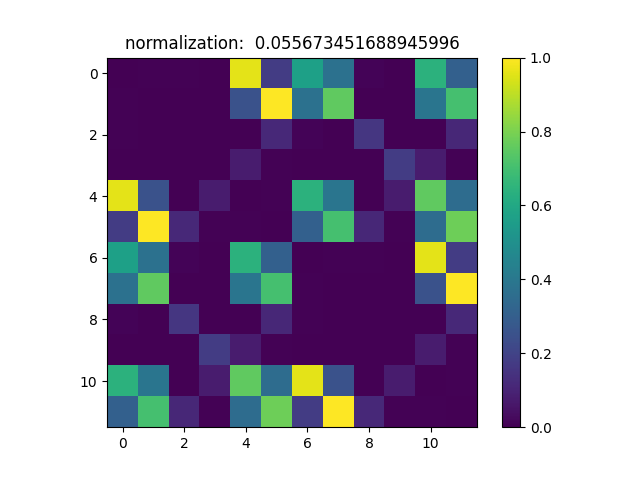}} \quad
\subfloat[Groundstate in natural orbitals]
   {\includegraphics[width=0.42\textwidth]{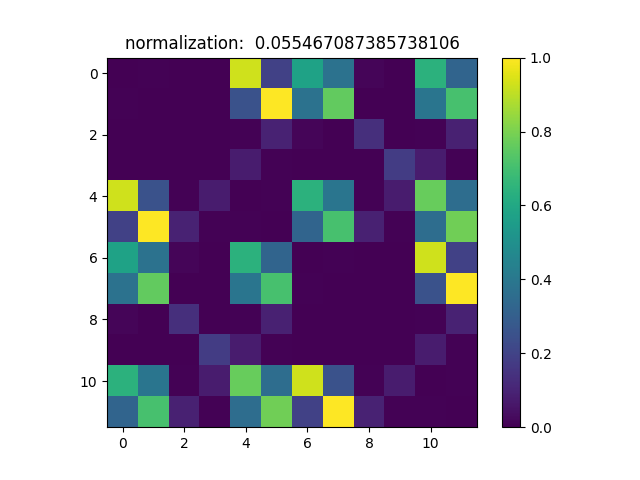}} \\
\caption{Quantum mutual information maps for $H_2O$ molecule. States and basis sets considered are described in the captions.}
\label{fig:H2O_WAHTOR}
\end{figure*}

In order to better understand and interpret the obtained result, let us proceed with a deeper analysis of the correlation between the occupation number of the spin-orbitals of the molecules. In the following, we will investigate the $H_2O$ molecule, albeit the same conclusions can be found when analysing the other systems, as illustrated in the supplemental material. 
We would like first to point out that, by using the Jordan-Wigner mapping, the occupancy of each spin-orbital is associated with the state of a single qubit. Therefore, on a given quantum state, the correlations between the occupancy of the spin-orbitals can be analyzed through the study of the correlations between the qubits and, for this purpose, we use the quantum mutual information matrix defined in the previous section. In particular, figure \ref{fig:H2O_WAHTOR} shows the quantum mutual information matrices for the states of the $H_2O$ molecule. 
Considering that the first half of the qubits represent the orbitals with spin up while the second half represents the same orbitals with spin down, each matrix is symmetrical and we can focus only on the lower or upper triangular part. 
We define the VQE and WAHTOR states as the ones obtained after the application of the corresponding algorithm. Then the VQE state is expressed in the Hartree-Fock molecular basis set and shown in panel (a) of the figure \ref{fig:H2O_WAHTOR} while the WAHTOR state, expressed in the WAHTOR-optimized single-particle basis set, is reported in panel (b). Panels (c), (d) and (e) show the quantum mutual information matrices for the ground state expressed respectively in the Hartree-Fock, WAHTOR-optimized and natural orbitals, respectively.
The colored spot ranges from 0 to 1 with the normalization constant reported above each figure.
First of all, we note that the maximal value of the correlation for image (b) is $33\%$ higher than the one in the image (a). 
Moreover, the number of green and yellow coloured spots in panel (b) is increased with respect to the ones in panel (a) while the number of blue spots decreases. This means that the quantum mutual information, during the optimization, has been compacted on a lower number of qubits pairs so that the sparsity of the quantum mutual information matrices is increased for the states expressed in the WAHTOR-optimized orbitals. This feature can be observed also comparing the ground state expressed in Hartree-Fock and WAHTOR-optimized molecular basis set reported in panels (c) and (d) respectively. Finally, we note that figures (d) and (e) are identical, confirming in terms of the correlation between the occupation number of spin-orbitals that the single-particle basis set obtained with the optimization process is almost identical to the natural one. 

\begin{figure}
\includegraphics[width=0.95\linewidth]{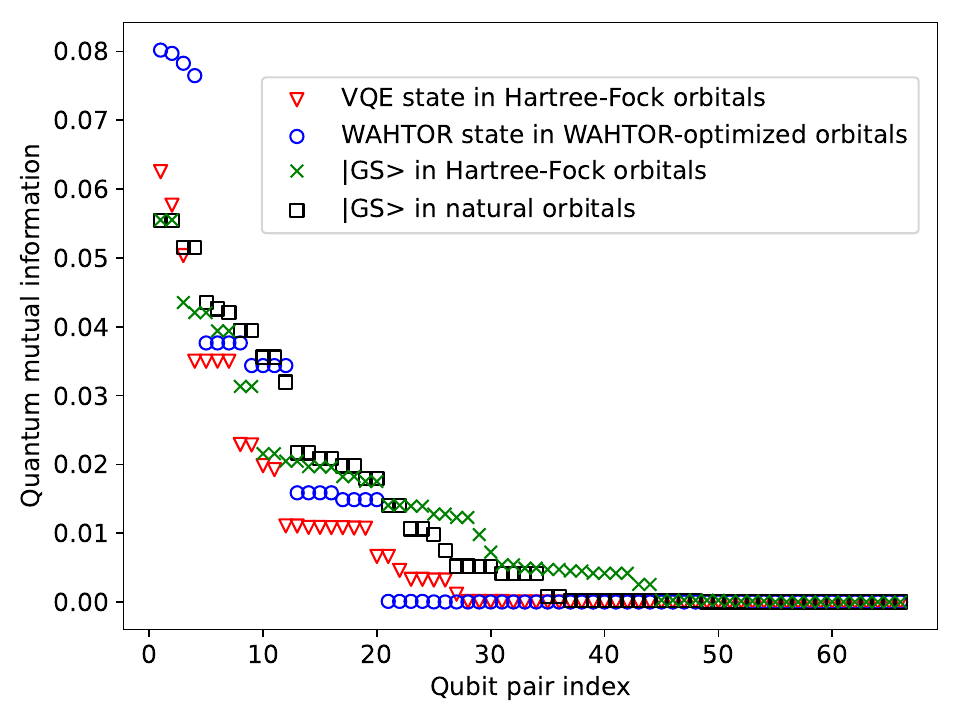}
\caption{Quantum mutual information values for qubit pairs in the $H_2O$ molecule are shown ordered in descending way. For the exact ground state results (green cross and black squares) is evident how the quantum mutual information is concentrated on a smaller number of qubit pairs in the natural orbital basis. The same finding is achieved for the variational solutions (red triangle and blue circles). In the case of WAHTOR optimised orbitals only about 20 qubit pairs have quantum mutual information values significantly different from zero.}
\label{fig:H2O_mutual_info}
\end{figure}

Figure \ref{fig:H2O_mutual_info} quantifies the variations in quantum mutual information between qubits pairs sorted in a decreasing way for the states represented in the matrices of figures \ref{fig:H2O_WAHTOR}(a), \ref{fig:H2O_WAHTOR}(b), \ref{fig:H2O_WAHTOR}(c) and \ref{fig:H2O_WAHTOR}(d).
Remarkably, the number of qubits pairs with values different from zero is smaller for the WAHTOR case than for the VQE state.
In the same manner, we observe that the number of uncorrelated qubit pairs is increased if the ground state is expressed in the WAHTOR-optimized orbitals rather than in the canonical Hartree-Fock molecular orbitals confirming the reduction of the sparsity in the matrices illustrated in figure \ref{fig:H2O_WAHTOR}.
Moreover, the maximal value of correlation between the qubit pairs is increased during the optimization passing from the state of figure \ref{fig:H2O_WAHTOR}(a) to the state of figure \ref{fig:H2O_WAHTOR}(b). 
The values of the quantum mutual information for each qubit pair are reported in figure \ref{fig:H2O_mutual_info} and ordered in descending way for both the exact and variational wavefunctions. We note that for the exact ground state the number of qubit pairs significantly different from zero is smaller when the wavefunction is expressed in natural orbitals with respect to the case of Hartree-Fock canonical orbitals. The same trend is observed for the variational wavefunctions.
In supplementary material, we show that there is a reduction of the sparsity in the matrices also when the other molecules shown in table \ref{table_molecule_results} are taken into account. Focusing on the converged molecular basis set, it has been conjectured in ref. \cite{Loewdin1955} that, for a given accuracy, the natural orbitals are the single-particle basis set in which the wavefunction is expressed with the lowest number of Slater determinants. In terms of correlations between qubits 
,we expect that as the number of determinants in the wavefunction increases, as more the number of coloured spots in the quantum mutual information matrices increases. This interpretation explains why the sparsity of the quantum mutual information matrices is increased when we express the states in the natural orbitals basis set. 
\section{CONCLUSIONS}
Natural orbitals diagonalize the one-particle reduced density matrix and it is supposed they are the orbitals in which the ground state is represented by a minimal number of Slater determinants. In this work we have investigated whether our orbital optimization procedure of an empirical ansatz for quantum chemistry is driving orbitals towards natural orbitals. For this purpose we have introduced a parameter $\delta$, as reported in table \ref{table_molecule_results} showing for different benchmark molecules that $\delta$ is very close to 1 in case of optimised WAHTOR orbitals. This convergence to natural orbitals is reached regardless of the geometric symmetries, the number of qubits and the accuracy achieved from the VQE results, for the molecules under consideration.

This effect is also reflected in terms of correlations: the comparison of the mutual information matrices built on Hartree-Fock and natural orbitals demonstrates that in the natural orbital basis there is an increase of the sparsity of the matrix as reported in figure \ref{fig:H2O_WAHTOR} for the case for the $H_2O$ molecule. Similar considerations can be deduced for all the other molecules by comparing the pictures in the supplemental material. 
The identification of natural orbitals as optimal orbitals for the building of empirical ans\"atze, may suggest them as being the best candidate one-electron basis for developing compact empirical wavefunction for state preparation on NISQ devices.

\section{Acknowledgments}
We thank Celestino Angeli for helpful discussions. We acknowledge fundings from the Marie Skłodowska-Curie Innovative Training Network (ITN) “MOQS - Molecular Quantum Simulations”, grant agreement number 955479; fundings from Ministero dell’Istruzione dell’Università e della Ricerca under project PON R \& I 2014-2020 and project PRIN 2022 number 2022W9W423.
\section{Competing interests}
The authors declare no competing financial interest.


\bibliographystyle{ieeetr} 
\bibliography{manuscript} 

\begin{thebibliography}{10}

\bibitem{Kitaev1995}
A.~Y. Kitaev, ``{Quantum measurements and the Abelian Stabilizer Problem},''
  pp.~1--22, 1995.

\bibitem{Abrams1997}
D.~S. Abrams and S.~Lloyd, ``{Simulation of Many-Body Fermi Systems on a
  Universal Quantum Computer},'' {\em Phys. Rev. Lett.}, vol.~79,
  pp.~2586--2589, sep 1997.

\bibitem{Abrams1999}
D.~S. Abrams and S.~Lloyd, ``{Quantum Algorithm Providing Exponential Speed
  Increase for Finding Eigenvalues and Eigenvectors},'' {\em Phys. Rev. Lett.},
  vol.~83, pp.~5162--5165, dec 1999.

\bibitem{Nielsen2010}
M.~A. Nielsen and I.~L. Chuang, {\em {Quantum Computation and Quantum
  Information}}.
\newblock Cambridge: Cambridge University Press, 2010.

\bibitem{Bauman2021}
N.~P. Bauman, H.~Liu, E.~J. Bylaska, S.~Krishnamoorthy, G.~H. Low, C.~E.
  Granade, N.~Wiebe, N.~A. Baker, B.~Peng, M.~Roetteler, M.~Troyer, and
  K.~Kowalski, ``{Toward Quantum Computing for High-Energy Excited States in
  Molecular Systems: Quantum Phase Estimations of Core-Level States},'' {\em J.
  Chem. Theory Comput.}, vol.~17, no.~1, pp.~201--210, 2021.

\bibitem{Barison2021}
S.~Barison, F.~Vicentini, and G.~Carleo, ``{An efficient quantum algorithm for
  the time evolution of parameterized circuits},'' {\em Quantum}, vol.~5,
  p.~512, jul 2021.

\bibitem{McArdle2019}
S.~McArdle, T.~Jones, S.~Endo, Y.~Li, S.~C. Benjamin, and X.~Yuan,
  ``{Variational ansatz-based quantum simulation of imaginary time
  evolution},'' {\em npj Quantum Inf.}, vol.~5, no.~1, pp.~1--14, 2019.

\bibitem{Cilimberto2018}
C.~Ciliberto, M.~Herbster, A.~D. Ialongo, M.~Pontil, A.~Rocchetto, S.~Severini,
  and L.~Wossnig, ``{Quantum machine learning: a classical perspective},'' {\em
  Proc. R. Soc. A Math. Phys. Eng. Sci.}, vol.~474, no.~2209, p.~20170551,
  2018.

\bibitem{Dunjko2020}
V.~Dunjko and P.~Wittek, ``{A non-review of {\{}Q{\}}uantum {\{}M{\}}achine
  {\{}L{\}}earning: trends and explorations},'' {\em Quantum Views}, vol.~4,
  p.~32, mar 2020.

\bibitem{Nannicini2019}
G.~Nannicini, ``{Performance of hybrid quantum-classical variational heuristics
  for combinatorial optimization},'' {\em Phys. Rev. E}, vol.~99, p.~13304, jan
  2019.

\bibitem{Egger2021}
D.~J. Egger, C.~Gambella, J.~Marecek, S.~McFaddin, M.~Mevissen, R.~Raymond,
  A.~Simonetto, S.~Woerner, and E.~Yndurain, ``{Quantum Computing for Finance:
  State-of-the-Art and Future Prospects},'' {\em IEEE Trans. Quantum Eng.},
  vol.~1, pp.~1--24, jan 2021.

\bibitem{Huggins2020}
W.~J. Huggins, J.~Lee, U.~Baek, B.~O'Gorman, and K.~B. Whaley, ``{A
  non-orthogonal variational quantum eigensolver},'' {\em New J. Phys.},
  vol.~22, no.~7, 2020.

\bibitem{OMalley2016}
P.~J. O'Malley, R.~Babbush, I.~D. Kivlichan, J.~Romero, J.~R. McClean,
  R.~Barends, J.~Kelly, P.~Roushan, A.~Tranter, N.~Ding, B.~Campbell, Y.~Chen,
  Z.~Chen, B.~Chiaro, A.~Dunsworth, A.~G. Fowler, E.~Jeffrey, E.~Lucero,
  A.~Megrant, J.~Y. Mutus, M.~Neeley, C.~Neill, C.~Quintana, D.~Sank,
  A.~Vainsencher, J.~Wenner, T.~C. White, P.~V. Coveney, P.~J. Love, H.~Neven,
  A.~Aspuru-Guzik, and J.~M. Martinis, ``{Scalable quantum simulation of
  molecular energies},'' {\em Phys. Rev. X}, vol.~6, no.~3, pp.~1--13, 2016.

\bibitem{Bauer2020}
B.~Bauer, S.~Bravyi, M.~Motta, and G.~K.-L. Chan, ``{Quantum Algorithms for
  Quantum Chemistry and Quantum Materials Science},'' {\em Chem. Rev.},
  vol.~120, pp.~12685--12717, nov 2020.

\bibitem{Fedorov2022}
D.~A. Fedorov, B.~Peng, N.~Govind, and Y.~Alexeev, ``{VQE method: a short
  survey and recent developments},'' {\em Mater. Theory}, vol.~6, p.~2, jan
  2022.

\bibitem{Cao2019}
Y.~Cao, J.~Romero, J.~P. Olson, M.~Degroote, P.~D. Johnson, M.~Kieferov{\'{a}},
  I.~D. Kivlichan, T.~Menke, B.~Peropadre, N.~P. Sawaya, S.~Sim, L.~Veis, and
  A.~Aspuru-Guzik, ``{Quantum Chemistry in the Age of Quantum Computing},''
  {\em Chem. Rev.}, vol.~119, no.~19, pp.~10856--10915, 2019.

\bibitem{Peruzzo2014}
A.~Peruzzo, J.~McClean, P.~Shadbolt, M.~H. Yung, X.~Q. Zhou, P.~J. Love,
  A.~Aspuru-Guzik, and J.~L. O'Brien, ``{A variational eigenvalue solver on a
  photonic quantum processor},'' {\em Nat. Commun.}, vol.~5, jul 2014.

\bibitem{Corcoles2020}
A.~D. Corcoles, A.~Kandala, A.~Javadi-Abhari, D.~T. McClure, A.~W. Cross,
  K.~Temme, P.~D. Nation, M.~Steffen, and J.~M. Gambetta, ``{Challenges and
  Opportunities of Near-Term Quantum Computing Systems},'' {\em Proc. IEEE},
  vol.~108, no.~8, pp.~1338--1352, 2020.

\bibitem{Preskill2018}
J.~Preskill, ``{Quantum computing in the NISQ era and beyond},'' {\em Quantum},
  vol.~2, no.~July, pp.~1--20, 2018.

\bibitem{Moll2018}
N.~Moll, P.~Barkoutsos, L.~S. Bishop, J.~M. Chow, A.~Cross, D.~J. Egger,
  S.~Filipp, A.~Fuhrer, J.~M. Gambetta, M.~Ganzhorn, A.~Kandala, A.~Mezzacapo,
  P.~M{\"{u}}ller, W.~Riess, G.~Salis, J.~Smolin, I.~Tavernelli, and K.~Temme,
  ``{Quantum optimization using variational algorithms on near-term quantum
  devices},'' {\em Quantum Sci. Technol.}, vol.~3, p.~030503, jul 2018.

\bibitem{Jordan1928}
P.~Jordan and E.~Wigner, ``{{\"{U}}ber das Paulische {\"{A}}quivalenzverbot},''
  vol.~1927, 1928.

\bibitem{Bravyi2002}
S.~B. Bravyi and A.~Y. Kitaev, ``{Fermionic quantum computation},'' {\em Ann.
  Phys. (N. Y).}, vol.~298, no.~1, pp.~210--226, 2002.

\bibitem{McArdle2020}
S.~McArdle, S.~Endo, A.~Aspuru-Guzik, S.~C. Benjamin, and X.~Yuan, ``{Quantum
  computational chemistry},'' {\em Rev. Mod. Phys.}, vol.~92, no.~1, pp.~1--59,
  2020.

\bibitem{Lee2019}
J.~Lee, W.~J. Huggins, M.~Head-Gordon, and K.~B. Whaley, ``{Generalized Unitary
  Coupled Cluster Wave functions for Quantum Computation},'' {\em J. Chem.
  Theory Comput.}, vol.~15, no.~1, pp.~311--324, 2019.

\bibitem{Romero2019}
J.~Romero, R.~Babbush, J.~R. McClean, C.~Hempel, P.~J. Love, and
  A.~Aspuru-Guzik, ``{Strategies for quantum computing molecular energies using
  the unitary coupled cluster ansatz},'' {\em Quantum Sci. Technol.}, vol.~4,
  no.~1, pp.~1--18, 2019.

\bibitem{Kandala2017}
A.~Kandala, A.~Mezzacapo, K.~Temme, M.~Takita, M.~Brink, J.~M. Chow, and J.~M.
  Gambetta, ``{Hardware-efficient variational quantum eigensolver for small
  molecules and quantum magnets},'' {\em Nature}, vol.~549, no.~7671,
  pp.~242--246, 2017.

\bibitem{Grimsley2019}
H.~R. Grimsley, S.~E. Economou, E.~Barnes, and N.~J. Mayhall, ``{An adaptive
  variational algorithm for exact molecular simulations on a quantum
  computer},'' {\em Nat. Commun.}, vol.~10, no.~1, 2019.

\bibitem{Benfenati2021}
F.~Benfenati, G.~Mazzola, C.~Capecci, P.~K. Barkoutsos, P.~J. Ollitrault,
  I.~Tavernelli, and L.~Guidoni, ``{Improved Accuracy on Noisy Devices by
  Nonunitary Variational Quantum Eigensolver for Chemistry Applications},''
  {\em J. Chem. Theory Comput.}, vol.~17, no.~7, pp.~3946--3954, 2021.

\bibitem{Stair2021}
N.~H. Stair and F.~A. Evangelista, ``{Simulating Many-Body Systems with a
  Projective Quantum Eigensolver},'' {\em PRX Quantum}, vol.~2, p.~030301, jul
  2021.

\bibitem{Ganzhorn2019}
M.~Ganzhorn, D.~J. Egger, P.~Barkoutsos, P.~Ollitrault, G.~Salis, N.~Moll,
  M.~Roth, A.~Fuhrer, P.~Mueller, S.~Woerner, I.~Tavernelli, and S.~Filipp,
  ``{Gate-Efficient Simulation of Molecular Eigenstates on a Quantum
  Computer},'' {\em Phys. Rev. Appl.}, vol.~11, no.~4, 2019.

\bibitem{Egger2023}
D.~J. Egger, C.~Capecci, B.~Pokharel, P.~K. Barkoutsos, L.~E. Fischer,
  L.~Guidoni, and I.~Tavernelli, ``{A study of the pulse-based variational
  quantum eigensolver on cross-resonance based hardware},'' pp.~1--13, 2023.

\bibitem{Tkachenko2020}
N.~V. Tkachenko, J.~Sud, Y.~Zhang, S.~Tretiak, P.~M. Anisimov, A.~T. Arrasmith,
  P.~J. Coles, L.~Cincio, and P.~A. Dub, ``{Correlation-Informed Permutation of
  Qubits for Reducing Ansatz Depth in the Variational Quantum Eigensolver},''
  {\em PRX Quantum}, vol.~2, no.~2, p.~1, 2021.

\bibitem{Meitei2021}
O.~R. Meitei, B.~T. Gard, G.~S. Barron, D.~P. Pappas, S.~E. Economou,
  E.~Barnes, and N.~J. Mayhall, ``{Gate-free state preparation for fast
  variational quantum eigensolver simulations},'' {\em npj Quantum Inf.},
  vol.~7, no.~1, pp.~1--11, 2021.

\bibitem{Ratini2022}
L.~Ratini, C.~Capecci, F.~Benfenati, and L.~Guidoni, ``{Wave Function Adapted
  Hamiltonians for Quantum Computing},'' {\em J. Chem. Theory Comput.},
  vol.~18, no.~2, pp.~899--909, 2022.

\bibitem{Ratini2023}
L.~Ratini, C.~Capecci, and L.~Guidoni, ``{Optimization strategies in WAHTOR
  algorithm for quantum computing empirical ansatz: a comparative study},''
  2023.

\bibitem{Loewdin1955}
P.-O. L{\"{o}}wdin, ``{Quantum Theory of Many-Particle Systems. I. Physical
  Interpretations by Means of Density Matrices, Natural Spin-Orbitals, and
  Convergence Problems in the Method of Configurational Interaction},'' {\em
  Phys. Rev.}, vol.~97, pp.~1474--1489, mar 1955.

\bibitem{Davidson1972}
E.~R. Davidson, ``{Properties and Uses of Natural Orbitals},'' {\em Rev. Mod.
  Phys.}, vol.~44, pp.~451--464, jul 1972.

\bibitem{Szabo2012}
{Szabo Attila} and {Ostlund Neil S.}, {\em {Modern Quantum Chemistry:
  Introduction to Advanced Electronic Structure Theory}}.
\newblock Dover Publications, 2012.

\bibitem{pyscf}
Q.~Sun, T.~C. Berkelbach, N.~S. Blunt, G.~H. Booth, S.~Guo, Z.~Li, J.~Liu,
  J.~D. McClain, E.~R. Sayfutyarova, S.~Sharma, S.~Wouters, and G.~K.-L. Chan,
  ``{PySCF: the Python-based simulations of chemistry framework},'' {\em WIREs
  Comput. Mol. Sci.}, vol.~8, no.~1, p.~e1340, 2018.

\bibitem{qiskit}
G.~Aleksandrowicz, T.~Alexander, P.~Barkoutsos, L.~Bello, Y.~Ben-Haim,
  D.~Bucher, F.~J. Cabrera-Hern{\'{a}}ndez, J.~Carballo-Franquis, A.~Chen,
  C.-F. Chen, J.~M. Chow, A.~D. C{\'{o}}rcoles-Gonzales, A.~J. Cross, A.~Cross,
  J.~Cruz-Benito, C.~Culver, S.~D. L.~P. Gonz{\'{a}}lez, E.~D.~L. Torre,
  D.~Ding, E.~Dumitrescu, I.~Duran, P.~Eendebak, M.~Everitt, I.~F. Sertage,
  A.~Frisch, A.~Fuhrer, J.~Gambetta, B.~G. Gago, J.~Gomez-Mosquera,
  D.~Greenberg, I.~Hamamura, V.~Havlicek, J.~Hellmers, {\L}.~Herok, H.~Horii,
  S.~Hu, T.~Imamichi, T.~Itoko, A.~Javadi-Abhari, N.~Kanazawa, A.~Karazeev,
  K.~Krsulich, P.~Liu, Y.~Luh, Y.~Maeng, M.~Marques, F.~J.
  Mart{\'{i}}n-Fern{\'{a}}ndez, D.~T. McClure, D.~McKay, S.~Meesala,
  A.~Mezzacapo, N.~Moll, D.~M. Rodr{\'{i}}guez, G.~Nannicini, P.~Nation,
  P.~Ollitrault, L.~J. O'Riordan, H.~Paik, J.~P{\'{e}}rez, A.~Phan, M.~Pistoia,
  V.~Prutyanov, M.~Reuter, J.~Rice, A.~R. Davila, R.~H.~P. Rudy, M.~Ryu,
  N.~Sathaye, C.~Schnabel, E.~Schoute, K.~Setia, Y.~Shi, A.~Silva, Y.~Siraichi,
  S.~Sivarajah, J.~A. Smolin, M.~Soeken, H.~Takahashi, I.~Tavernelli,
  C.~Taylor, P.~Taylour, K.~Trabing, M.~Treinish, W.~Turner, D.~Vogt-Lee,
  C.~Vuillot, J.~A. Wildstrom, J.~Wilson, E.~Winston, C.~Wood, S.~Wood,
  S.~W{\"{o}}rner, I.~Y. Akhalwaya, and C.~Zoufal, ``{Qiskit: An Open-source
  Framework for Quantum Computing},'' jan 2019.

\bibitem{Byrd1995}
R.~H. Byrd, P.~Lu, J.~Nocedal, and C.~Zhu, ``{A limited memory algorithm for
  bound constrained optimization},'' {\em SIAM J. Sci. Comput.}, vol.~16,
  pp.~1190--1208, 1995.

\bibitem{Zeng2019}
B.~Zeng, X.~Chen, D.-L. Zhou, and X.-G. Wen, {\em {Quantum Information Meets
  Quantum Matter}}.
\newblock New York, NY: Springer New York, 2019.

\bibitem{Helgaker2000}
T.~Helgaker, P.~J{\o}rgensen, and J.~Olsen, {\em {Molecular Electronic
  Structure Theory}}.
\newblock Chichester: John Wiley {\&} Sons, LTD, 2000.

\bibitem{Giuliani2005}
G.~Giuliani and G.~Vignale, {\em {Quantum Theory of the Electron Liquid}}.
\newblock Cambridge University Press, mar 2005.

\end{thebibliography}

\newpage
\twocolumn[
\begin{@twocolumnfalse}
\section*
{\centering{Supplementary Material: Natural orbitals and sparsity of quantum mutual information, L. Ratini,C. Capecci and L. Guidoni}}
\label{supplemental}
Below are reported the quantum mutual information maps for the other molecules considered in this work. For each figure, we have the VQE state (a), the WAHTOR state (b) and the ground state (c,d,e). The first one is expressed in the HF basis, the second one in the converged molecular basis and the third one in three different basis sets: HF, converged and natural orbitals set. 
\end{@twocolumnfalse}]


\begin{figure*}[!htb]
\centering
\subfloat[VQE state in HF canonical orbitals]
   {\includegraphics[width=0.45\textwidth]{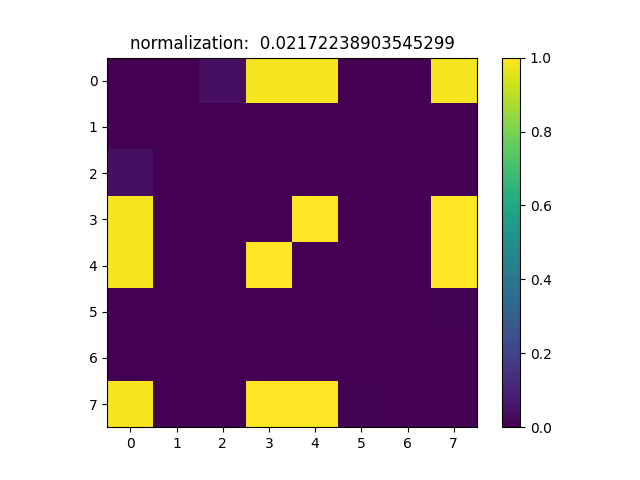}} \quad
\subfloat[WAHTOR state in WAHTOR-optimized orbitals]
   {\includegraphics[width=0.45\textwidth]{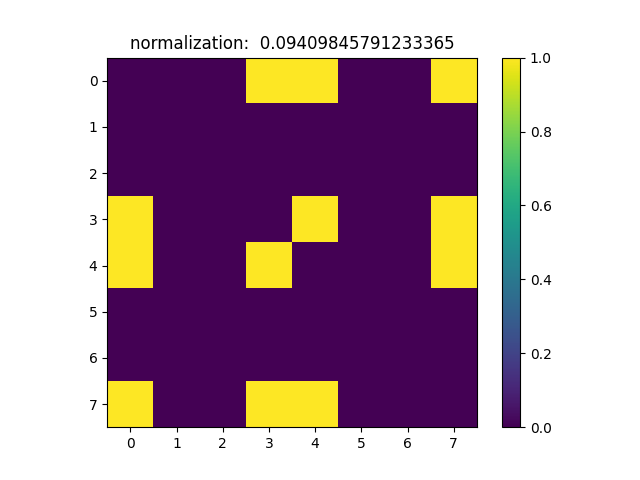}} \\
\subfloat[Groundstate in HF canonical orbitals]
   {\includegraphics[width=0.45\textwidth]{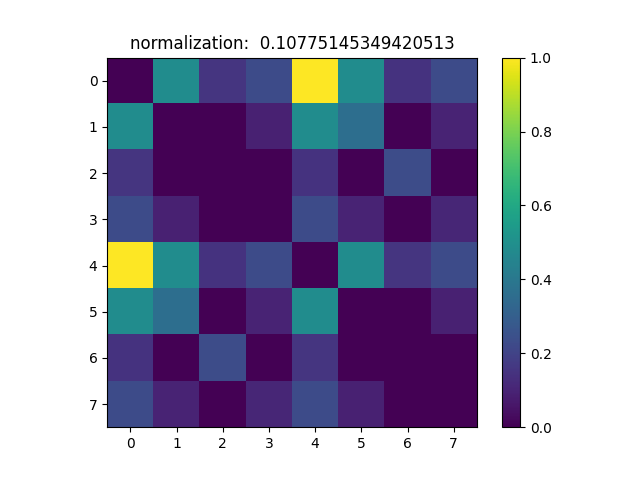}} \quad
\subfloat[Groundstate in WAHTOR-optimized orbitals]
   {\includegraphics[width=0.45\textwidth]{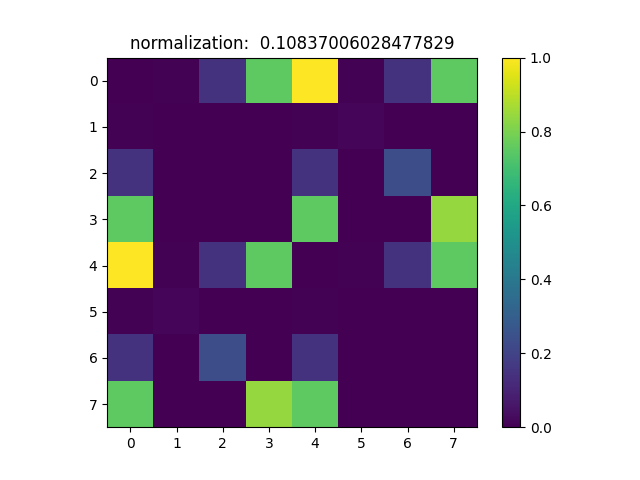}} \quad   
\subfloat[Groundstate in natural orbitals]
   {\includegraphics[width=0.45\textwidth]{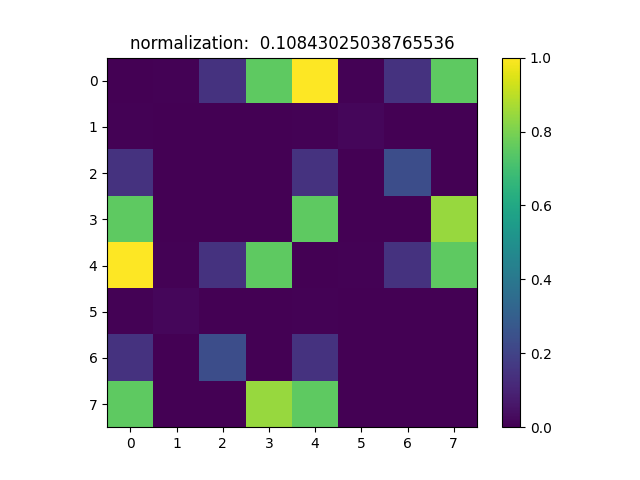}} \\
\caption{Quantum mutual information for $H_2$ molecule.}
\label{fig:H2}
\end{figure*}

\begin{figure*}[!htb]
\centering
\subfloat[VQE state in HF canonical orbitals]
   {\includegraphics[width=0.45\textwidth]{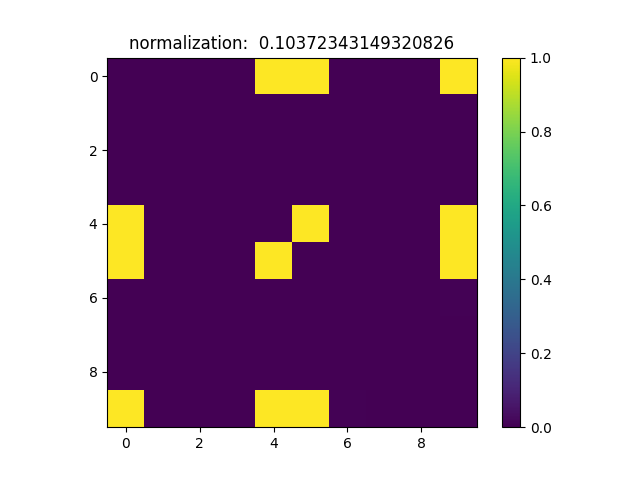}} \quad
\subfloat[WAHTOR state in WAHTOR-optimized orbitals]
   {\includegraphics[width=0.45\textwidth]{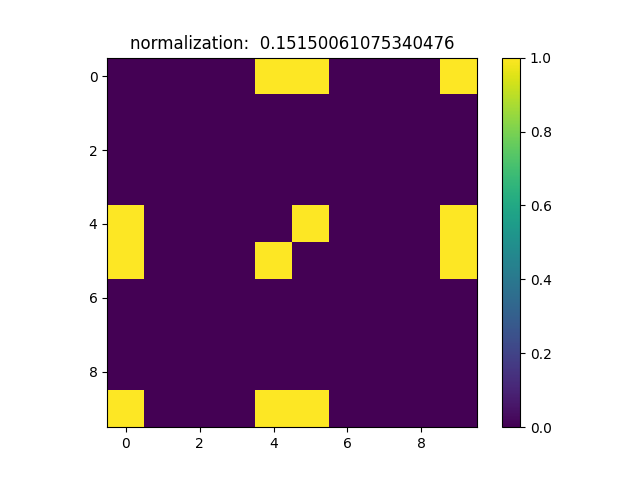}} \\
\subfloat[Groundstate in HF canonical orbitals]
   {\includegraphics[width=0.45\textwidth]{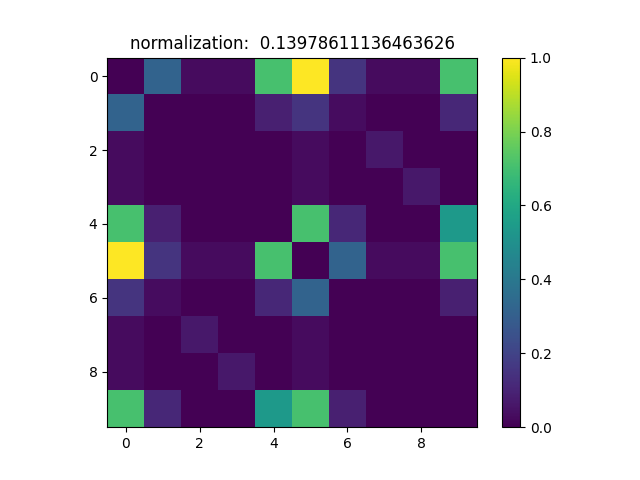}} \quad
\subfloat[Groundstate in WAHTOR-optimized orbitals]
   {\includegraphics[width=0.45\textwidth]{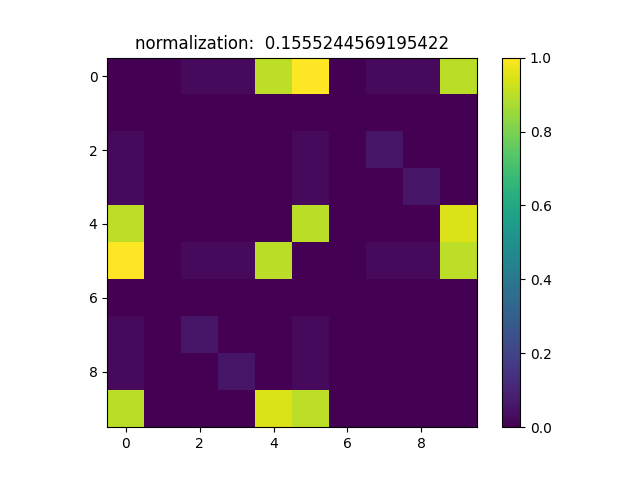}} \quad   
\subfloat[Groundstate in natural orbitals]
   {\includegraphics[width=0.45\textwidth]{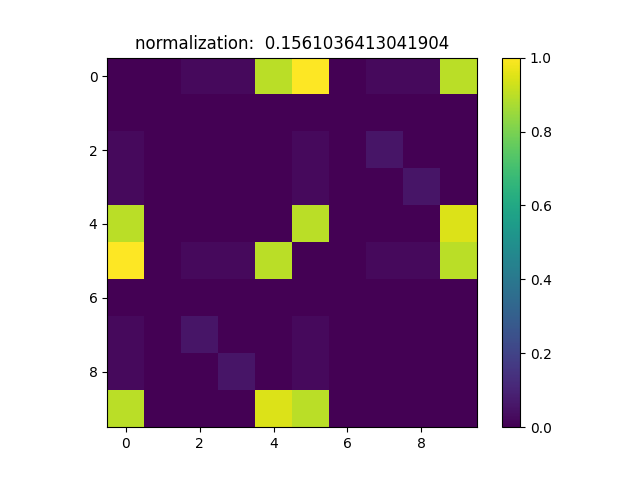}} \\
\caption{Quantum mutual information for $LiH$ molecule.}
\label{fig:LiH}
\end{figure*}

\begin{figure*}[!htb]
\centering
\subfloat[VQE state in HF canonical orbitals]
   {\includegraphics[width=0.45\textwidth]{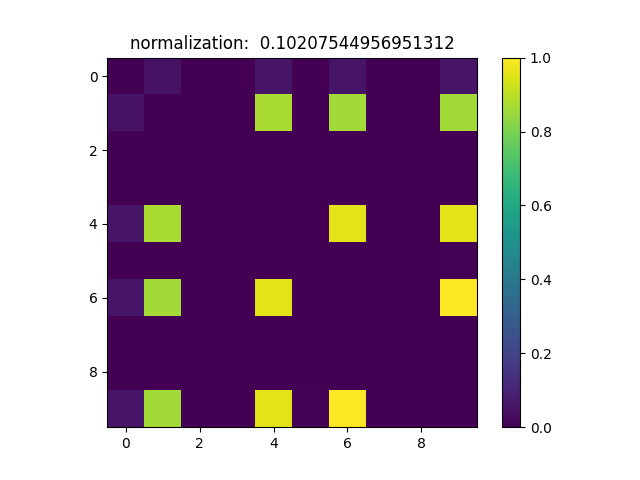}} \quad
\subfloat[WAHTOR state in WAHTOR-optimized orbitals]
   {\includegraphics[width=0.45\textwidth]{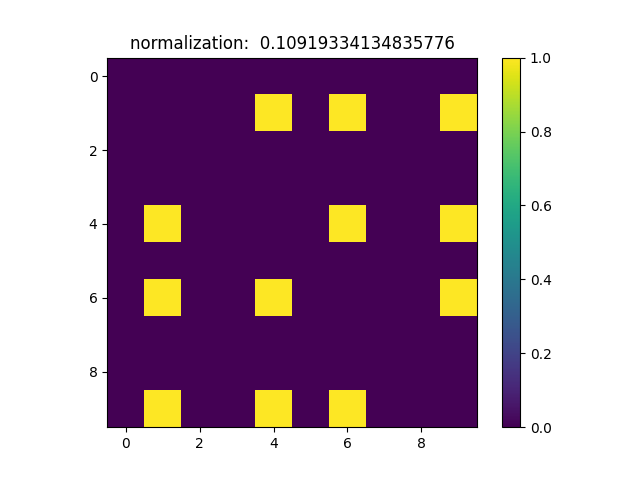}} \\
\subfloat[Groundstate in HF canonical orbitals]
   {\includegraphics[width=0.45\textwidth]{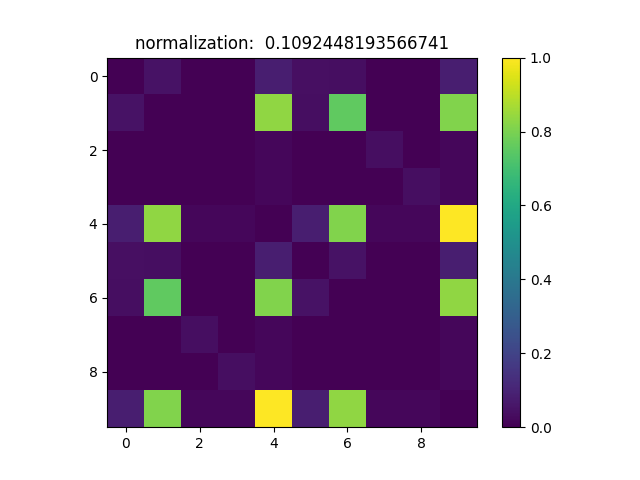}} \quad
\subfloat[Groundstate in WAHTOR-optimized orbitals]
   {\includegraphics[width=0.45\textwidth]{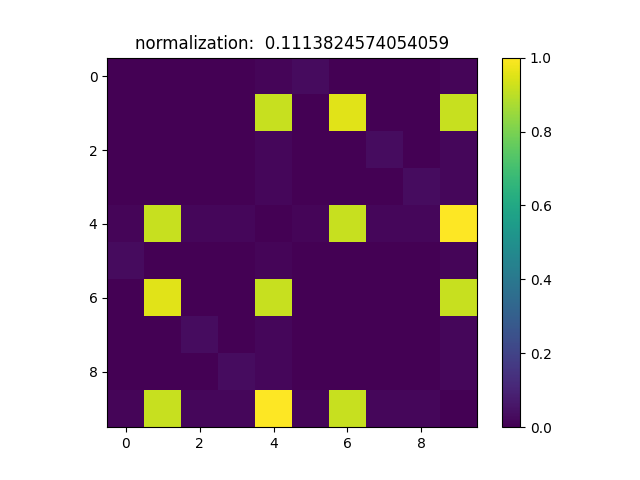}} \quad   
\subfloat[Groundstate in natural orbitals]
   {\includegraphics[width=0.45\textwidth]{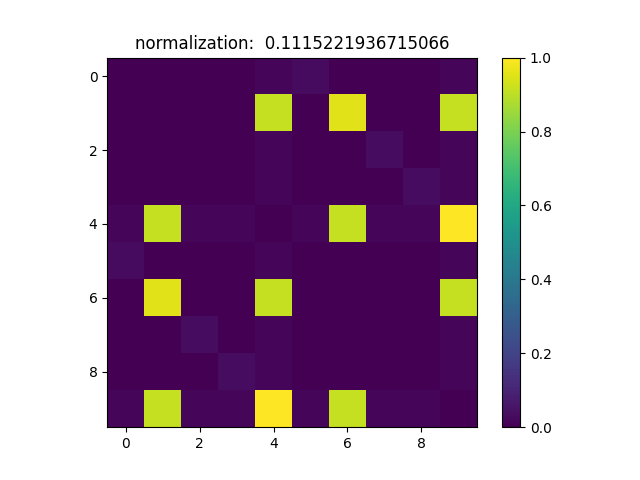}} \\
\caption{Quantum mutual information for $HF$ molecule.}
\label{fig:HF}
\end{figure*} 

\begin{figure*}[!htb]
\centering
\subfloat[VQE state in HF canonical orbitals]
   {\includegraphics[width=0.45\textwidth]{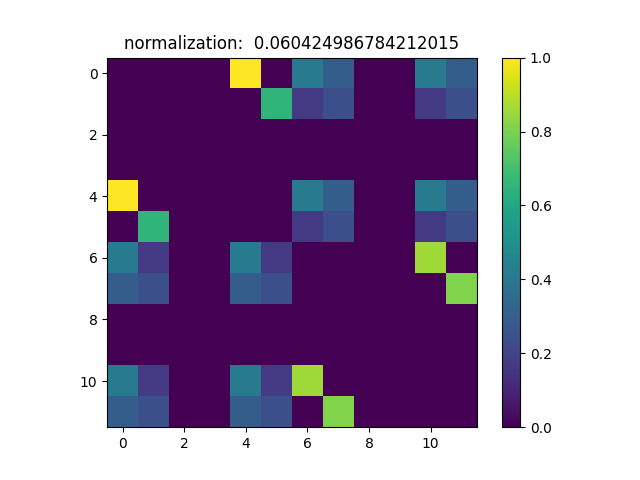}} \quad
\subfloat[WAHTOR state in WAHTOR-optimized orbitals]
   {\includegraphics[width=0.45\textwidth]{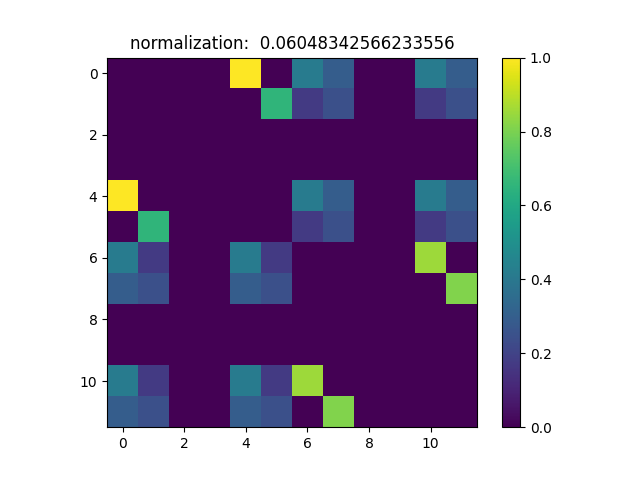}} \\
\subfloat[Groundstate in HF canonical orbitals]
   {\includegraphics[width=0.45\textwidth]{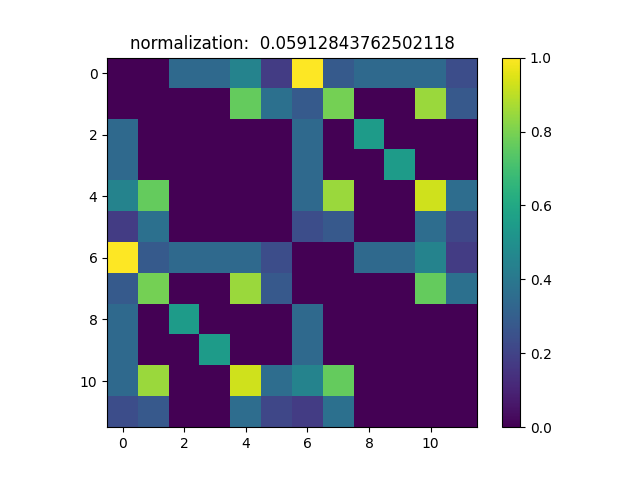}} \quad
\subfloat[Groundstate in WAHTOR-optimized orbitals]
   {\includegraphics[width=0.45\textwidth]{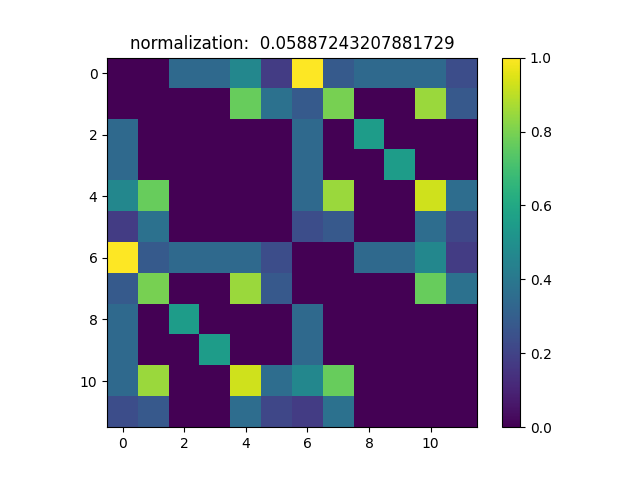}} \quad   
\subfloat[Groundstate in natural orbitals]
   {\includegraphics[width=0.45\textwidth]{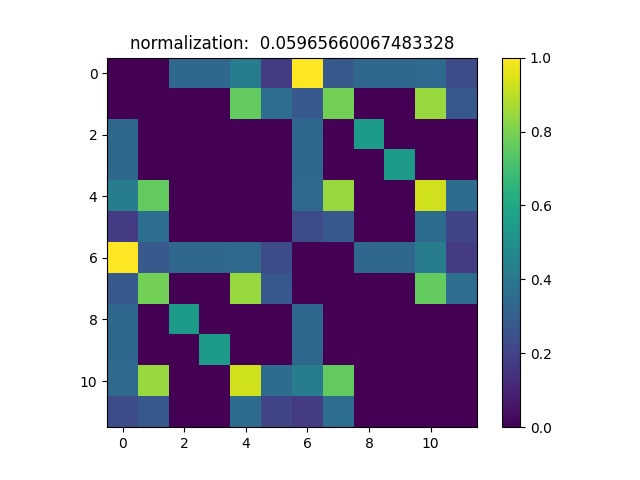}} \\
\caption{Quantum mutual information for $BeH_2$ molecule (case 1).}
\label{fig:BeH2_1}
\end{figure*} 

\begin{figure*}[!htb]
\centering
\subfloat[VQE state in HF canonical orbitals]
   {\includegraphics[width=0.45\textwidth]{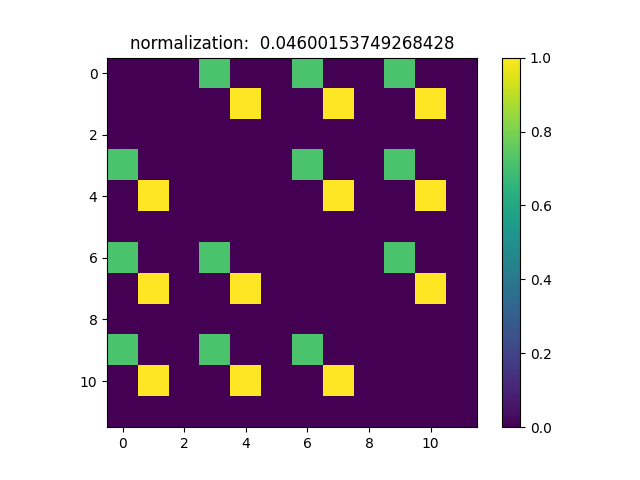}} \quad
\subfloat[WAHTOR state in WAHTOR-optimized orbitals]
   {\includegraphics[width=0.45\textwidth]{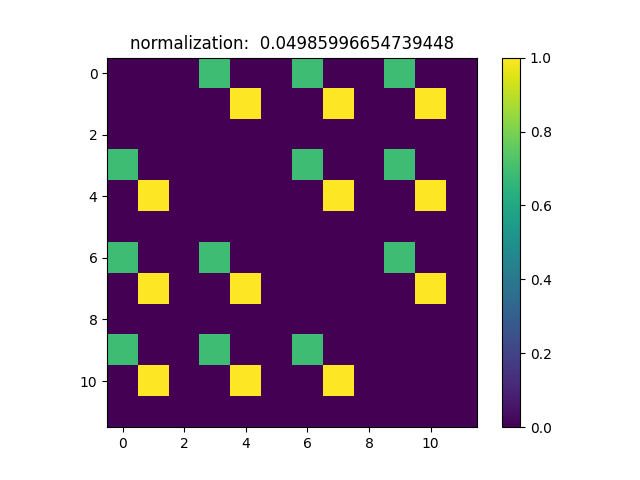}} \\
\subfloat[Groundstate in HF canonical orbitals]
   {\includegraphics[width=0.45\textwidth]{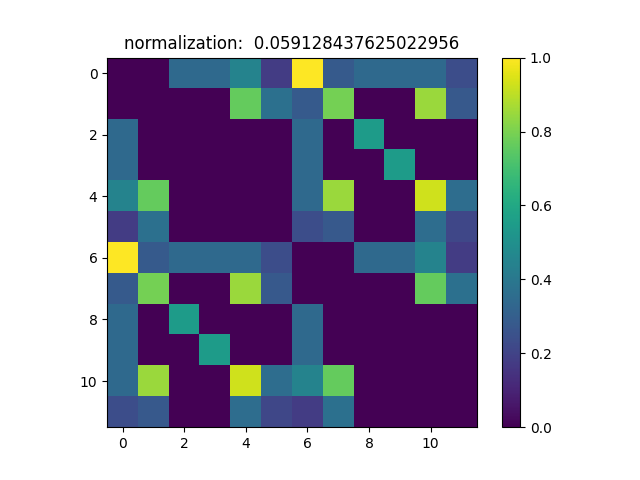}} \quad
\subfloat[Groundstate in WAHTOR-optimized orbitals]
   {\includegraphics[width=0.45\textwidth]{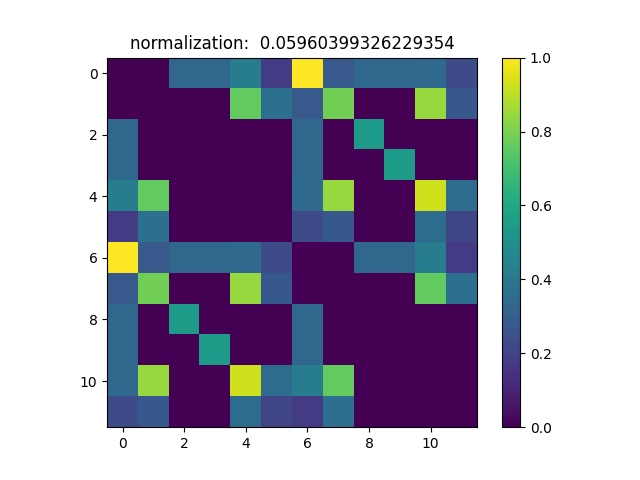}} \quad   
\subfloat[Groundstate in natural orbitals]
   {\includegraphics[width=0.45\textwidth]{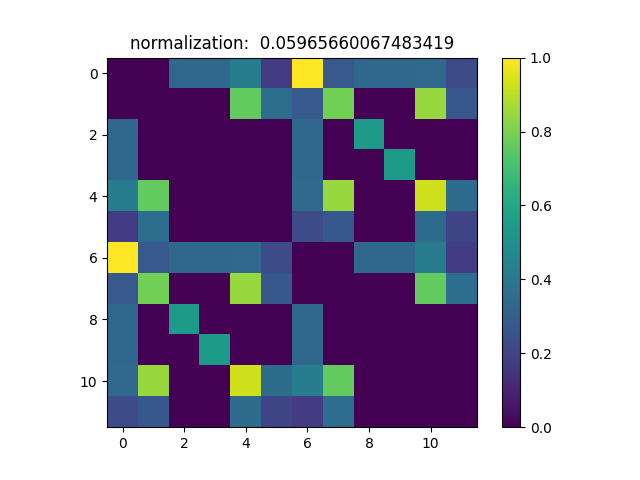}} \\
\caption{Quantum mutual information for $BeH_2$ molecule (case 2).}
\label{fig:BeH2_2}
\end{figure*} 

\begin{figure*}[!htb]
\centering
\subfloat[VQE state in HF canonical orbitals]
   {\includegraphics[width=0.45\textwidth]{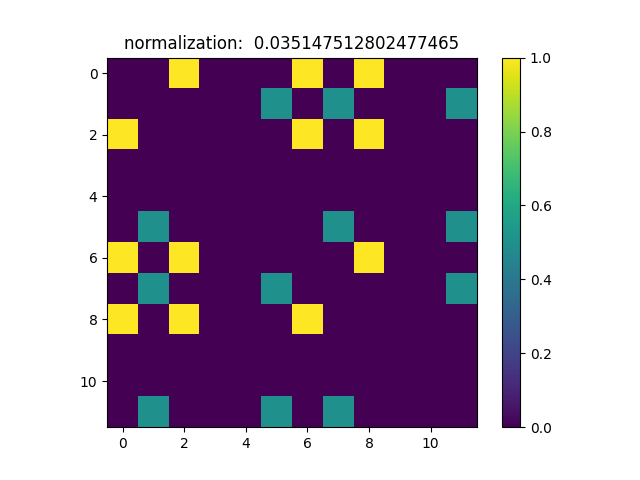}} \quad
\subfloat[WAHTOR state in WAHTOR-optimized orbitals]
   {\includegraphics[width=0.45\textwidth]{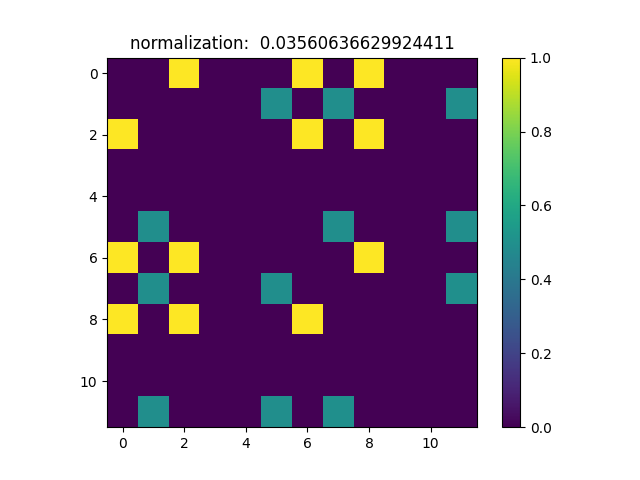}} \\
\subfloat[Groundstate in HF canonical orbitals]
   {\includegraphics[width=0.45\textwidth]{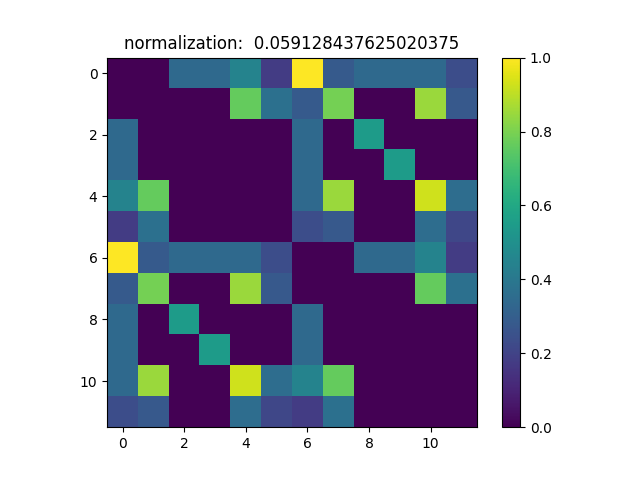}} \quad
\subfloat[Groundstate in WAHTOR-optimized orbitals]
   {\includegraphics[width=0.45\textwidth]{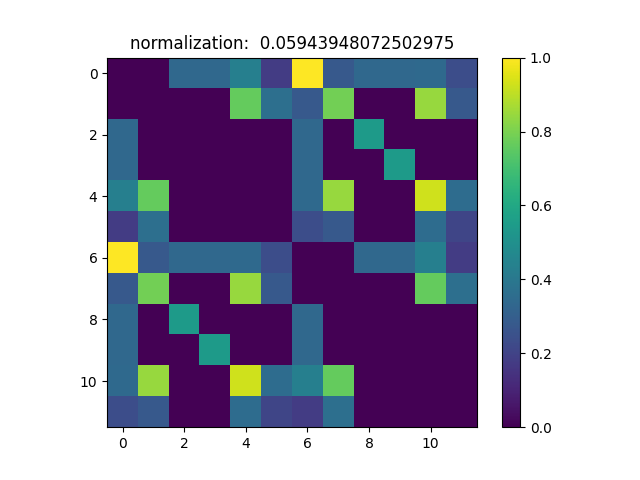}} \quad   
\subfloat[Groundstate in natural orbitals]
   {\includegraphics[width=0.45\textwidth]{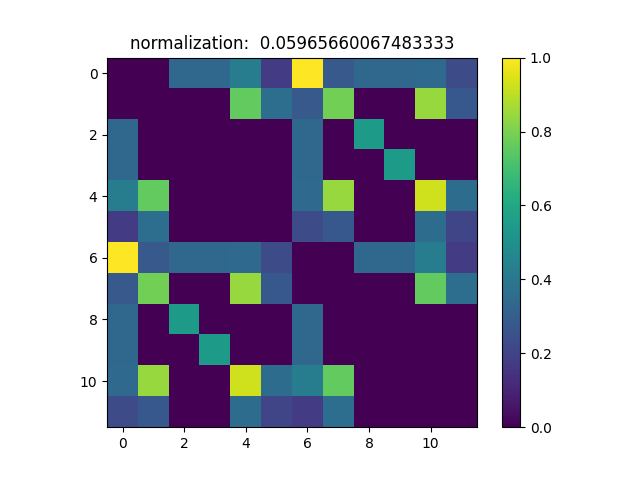}} \\
\caption{Quantum mutual information for $BeH_2$ molecule (case 3).}
\label{fig:BeH2_3}
\end{figure*} 

\begin{figure*}[!htb]
\centering
\subfloat[VQE state in HF canonical orbitals]
   {\includegraphics[width=0.45\textwidth]{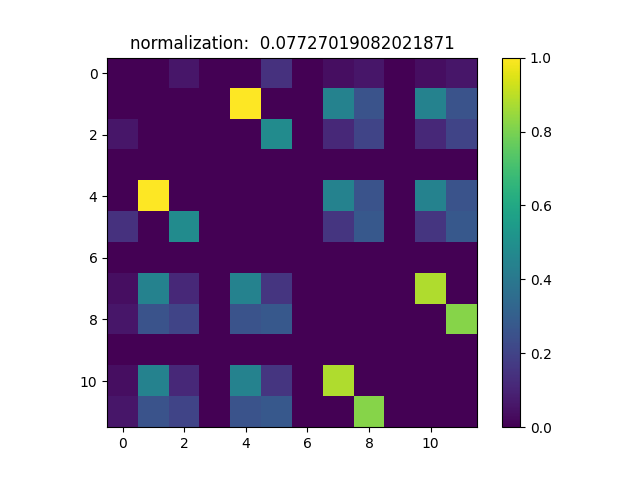}} \quad
\subfloat[WAHTOR state in WAHTOR-optimized orbitals]
   {\includegraphics[width=0.45\textwidth]{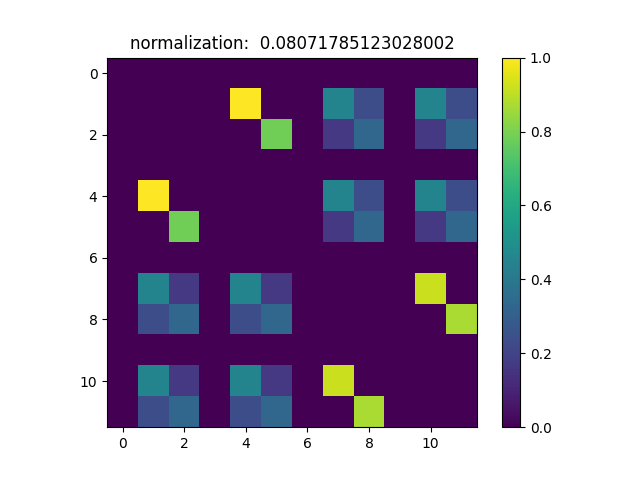}} \\
\subfloat[Groundstate in HF canonical orbitals]
   {\includegraphics[width=0.45\textwidth]{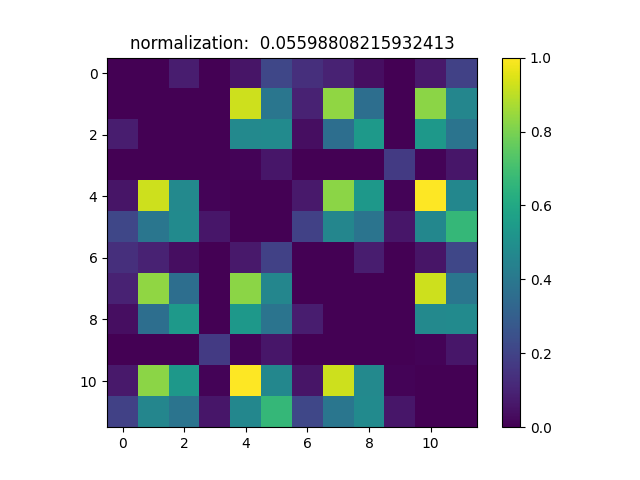}} \quad
\subfloat[Groundstate in WAHTOR-optimized orbitals]
   {\includegraphics[width=0.45\textwidth]{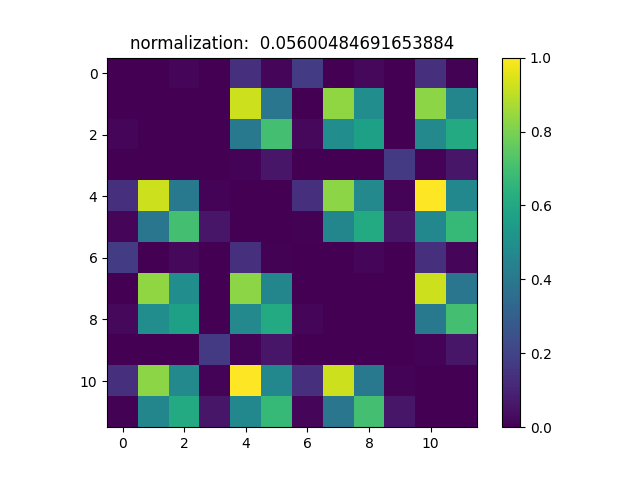}} \quad   
\subfloat[Groundstate in natural orbitals]
   {\includegraphics[width=0.45\textwidth]{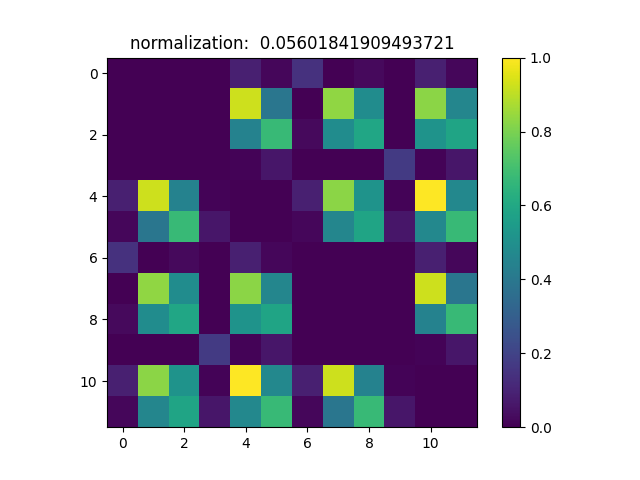}} \\
\caption{Quantum mutual information for $H_2S$ molecule.}
\label{fig:H2S}
\end{figure*} 

\begin{figure*}[!htb]
\centering
\subfloat[VQE state in HF canonical orbitals]
   {\includegraphics[width=0.45\textwidth]{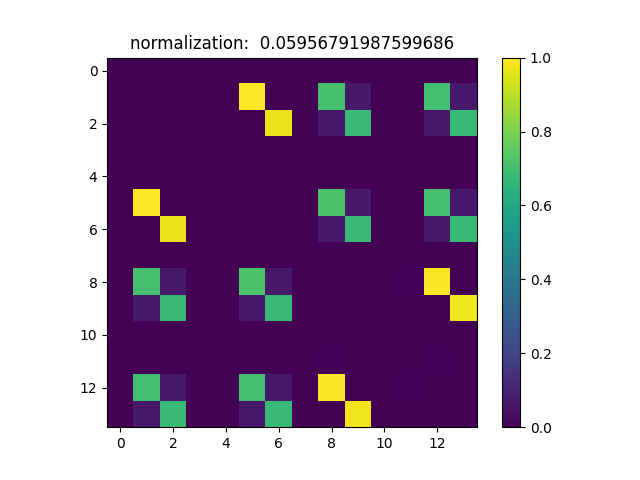}} \quad
\subfloat[WAHTOR state in WAHTOR-optimized orbitals]
   {\includegraphics[width=0.45\textwidth]{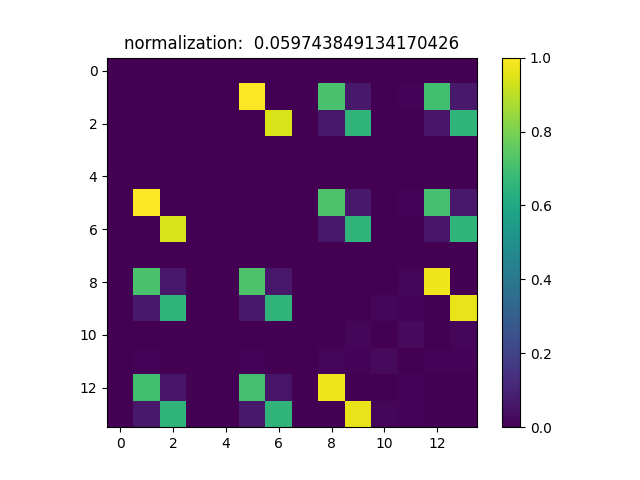}} \\
\subfloat[Groundstate in HF canonical orbitals]
   {\includegraphics[width=0.45\textwidth]{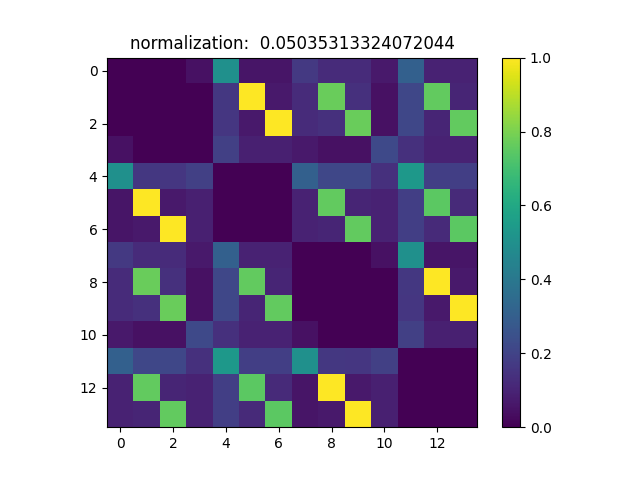}} \quad
\subfloat[Groundstate in WAHTOR-optimized orbitals]
   {\includegraphics[width=0.45\textwidth]{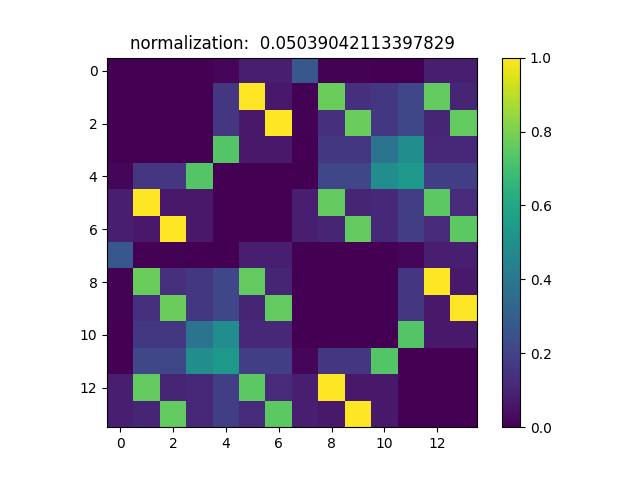}} \quad   
\subfloat[Groundstate in natural orbitals]
   {\includegraphics[width=0.45\textwidth]{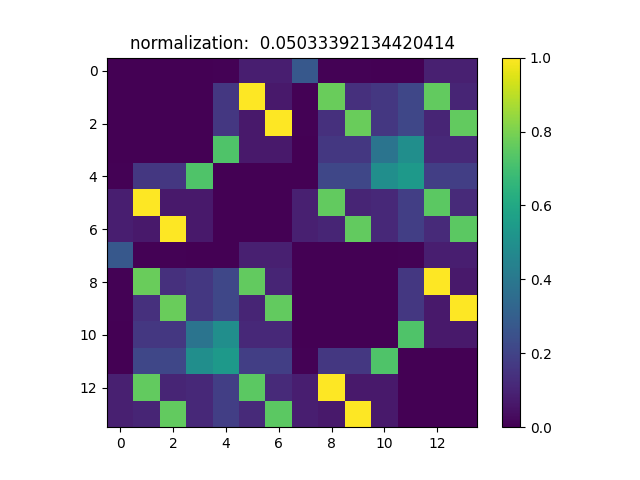}} \\
\caption{Quantum mutual information for $NH_3$ molecule.}
\label{fig:NH3}
\end{figure*}

\end{document}